\documentclass[twocolumn,longbibliography,floatfix,a4paper,accepted=2021-03-14]{quantumarticle}
\pdfoutput=1
\usepackage[utf8]{inputenc}
\usepackage[english]{babel}
\usepackage[T1]{fontenc}
\usepackage{amsmath}
\usepackage{hyperref}
\usepackage{amssymb}
\usepackage{amsfonts}
\newcommand{\vecg}{\boldsymbol}
\renewcommand{\vec}{\textbf}
\newcommand{\ket}[1]{|#1\rangle}
\newcommand{\bra}[1]{\langle#1|}
\newcommand{\bracket}[2]{\langle#1|#2\rangle}

\newcommand{\sqm}{\Omega}
\newcommand{\sqp}{\Omega}

\DeclareMathOperator{\tr}{Tr} 
\usepackage{subfigure}
\usepackage{graphicx}
\usepackage{color}

\usepackage[numbers,sort&compress]{natbib}

\def\bl{\begin{list}{}{\setlength{\partopsep}{0pt}\setlength{\itemsep}{0pt}%
			\setlength{\topsep}{0pt}\setlength{\parskip}{0pt}}}
\def\el{\end{list}\setlength{\parskip}{0pt}}

\begin{document}

\bibliographystyle{plainnat}
\title{Nonlinear extension of the quantum dynamical semigroup}

\author{Jakub Rembieli\'nski}
\email{jaremb@uni.lodz.pl}
\author{Pawe{\l}{} Caban}
\email{P.Caban@merlin.phys.uni.lodz.pl}

\affiliation{Department of Theoretical Physics,\\
	Faculty of Physics and Applied Informatics, University of Lodz\\
Pomorska 149/153, 90-236 {\L}{\'o}d{\'z}, Poland}

\begin{abstract}
In this paper we consider deterministic nonlinear time evolutions
satisfying so called convex quasi-linearity condition.
Such evolutions preserve the equivalence of ensembles and
therefore are free from problems with signaling.
We show that if family of linear
non-trace-preserving maps satisfies the semigroup property then the
generated family of convex quasi-linear operations also possesses 
the semigroup property. Next we generalize the
Gorini--Kossakowski--Sudarshan--Lindblad type equation for
the considered evolution.
As examples we discuss the 
general qubit evolution
in our model as well as an extension of the
Jaynes--Cummings model.
We apply our formalism to spin density matrix of a charged 
particle moving in the electromagnetic field as well as to flavor 
evolution of  solar neutrinos.
\end{abstract}
\maketitle

\section{Introduction}

In last decades many authors tried to generalize the standard quantum
mechanical evolution. Two most important approaches are based on
including nonlinear operations (see, e.g., \cite{BM1976_Nonlinear,Weinberg1989_Nonlin_Annals,Weinberg1989_PRL_nonlin})
and considering non-Hermitian Hamiltonians (see, e.g., \cite{Bender2007_Non-Hermitian,Moiseyev2011}).

Deterministic nonlinear evolutions are believed to allow for signaling
\cite{GR1995_relevant_Sch}, which was for the first time explicitly shown
by Gisin in \cite{Gisin1990_Weinberg_nonlin}, compare also
\cite{Polchinski1991_on_Weinberg,Czachor1991}.
Such evolutions are usually defined for pure states:
\begin{equation}
f_t\colon \ket{\psi}\to\ket{\psi(t)},
\end{equation}
and consequently the evolution of ensembles is assumed to have
the following form:
If $f_t(\ket{\psi_1})=\ket{\psi_1(t)}$,
$f_t(\ket{\psi_2})=\ket{\psi_2(t)}$ then
\begin{multline}
f_t\colon \lambda \ket{\psi_1}\bra{\psi_1}+
(1-\lambda) \ket{\psi_2}\bra{\psi_2} \to\\
\lambda \ket{\psi_1(t)}\bra{\psi_1(t)}+
(1-\lambda) \ket{\psi_2(t)}\bra{\psi_2(t)},
\label{evol_ensembles_standard}
\end{multline}
i.e., coefficients of the ensemble do not change under the evolution.
This assumption easily implies that deterministic nonlinear evolution
breaks the equivalence of ensembles corresponding to the same
mixed state and results in the possibility of arbitrary fast signaling
\cite{BK2015_Faster-then-light-linear}. In our recent paper
\cite{RC2019_Nonlinearity_signaling} it was shown that if we replace 
the assumption (\ref{evol_ensembles_standard}) by the 
so called convex quasi-linearity, i.e., we allow the appropriate
change of the coefficients
[see Eq.~(\ref{convex_quasi-linear})], then evolution satisfying
such a condition preserves equivalence of ensembles and 
consequently such evolutions cannot be ruled out by the 
standard Gisin's argument.

Let us stress that in our approach we do not change anything else
in the quantum formalism but only extend admissible set of
quantum evolutions by including nonlinear deterministic evolutions
that do not admit superluminal signaling. 
This is in contrast with such
nonlinear extensions of quantum mechanics that does not 
allow
signaling but demand modifications of other quantum mechanical rules
(see, e.g., \cite{CD2002_Correl-exp-nonlin_PLA,HCh2017_Born_NLQM}).
We also do not consider here stochastic nonlinear evolutions some of which
are free from the problems with signaling and have many important
applications (like in the collapse models \cite{GRW1986,BLSSU2013}).

In this paper we demonstrate that there exist a large class of 
convex quasi-linear evolutions. 
These evolutions are generated by linear non-trace-preserving
quantum operations
and/or derived from generalized master equation. 
What is interesting, recently considered evolutions
generated by non-Hermitian Hamiltonians \cite{SZ2013_non-Hermitian_q_dyn_two-level,GCKM_su11_Hamiltonian,%
BG2012_non-Hermitian-gain-loss,KAU2017_Non-Hermitian-information-retrieval}
also belong to this class.
It shows that convex quasi-linearity might be used as a principle joining
nonlinear quantum mechanics and non-Hermitian quantum mechanics.

In Sec.~\ref{sec:nonlin-q-operations} we remind the definition of 
convex quasi-linearity 
formulated in \cite{RC2019_Nonlinearity_signaling} 
and demonstrate that each 
linear non-trace-preserving
quantum operation generates convex quasi-linear operation.
In Sec.~\ref{sec:nonlin_evolutions} we show that if family of linear
non-trace-preserving maps satisfies the semigroup property then the
generated family of convex quasi-linear operations also possesses 
the semigroup property. Next we consider 
Gorini--Kossakowski--Sudarshan--Lindblad type equation for
the considered evolution.
Sec.~\ref{sec:qubit} is devoted to the discussion of qubit evolution
in our model
while in Sec.~\ref{sec:applications} we apply our 
formalism to two physical systems.
We finish with some discussion and  conclusions.

\section{Admissible nonlinear quantum operations}
\label{sec:nonlin-q-operations}

We start with recalling the definition of convex quasi-linearity
\cite{RC2019_Nonlinearity_signaling}.
Let us denote by $S$ the convex set of density matrices.
We call a map $\Phi\colon S\to S$ convex quasi-linear if
for all $\rho_i\in S$ and $p_i\in\langle0,1\rangle$, 
$\sum_i p_i=1$ ($i=1,\dots, N$ ) there exist $\bar{p}_i$ such that
\begin{equation}
\Phi\big[\sum_i p_i \rho_i\big] = \sum_i \bar{p}_i \Phi[\rho_i]
\label{convex_quasi-linear}
\end{equation}
and $\bar{p}_i\in\langle0,1\rangle$, 
$\sum_i \bar{p}_i=1$.
Below we show that there exists a class of convex quasi-linear
operations.

An example is discussed by Kraus  \cite{Kraus1983} as a generalized
measurement. We considered
this example in details in our previous paper \cite{RC2019_Nonlinearity_signaling}.

Let us consider the most general linear quantum operation
$\varphi$ having the Kraus form
\begin{equation}
\rho_{in}\mapsto 
\varphi(\rho_{in}) = \sum_{\alpha=0}^{\alpha_{max}}
K_\alpha \rho_{in} K_{\alpha}^{\dagger},
\end{equation}
where $\tr(\rho_{in})=1$, $K_\alpha$ are Kraus operators, 
$\alpha_{max}< N^2$, $N<\infty$ is the  dimension of the Hilbert
space $\mathcal{H}$ of the considered system.
Furthermore,   $F=\sum_{\alpha} K_{\alpha}^{\dagger} K_\alpha \le I$.
To obtain a map into the convex set of density matrices we must normalize
the $\varphi(\rho_{in})$ in the case $F<I$.
Consequently, the  complete quantum operation has the form
\begin{align}
\rho_{out} = \Phi(\rho_{in}) & = \frac{1}{\tr(\sum_{\beta} K_\beta \rho_{in} K_{\beta}^{\dagger})} 
\sum_\alpha K_\alpha \rho_{in} K_{\alpha}^{\dagger}\\
& = \frac{1}{\tr(F \rho_{in})}
\sum_\alpha K_\alpha \rho_{in} K_{\alpha}^{\dagger},
\end{align}
i.e., 
\begin{equation}
\Phi(\rho_{in}) = \frac{\varphi(\rho_{in})}{\tr[\varphi(\rho_{in})]}.
\end{equation}
Of course $\tr[\varphi(\rho_{in})]=1$ implies linearity of $\Phi$.
It is easy to see that $\Phi$ is in general convex  quasi-linear, i.e.,
\begin{equation}
\Phi(\lambda \rho_{in}^{a} + (1-\lambda)\rho_{in}^{b})
= \bar{\lambda} \Phi(\rho_{in}^{a}) 
+ (1-\bar{\lambda}) \Phi(\rho_{in}^{b}),
\end{equation}
where $0\le\lambda\le 1$ and thus
\begin{align}
\bar{\lambda}& =\lambda\frac{\tr(F\rho_{in}^{a})}{\tr(F\rho_{in})}\ge0\\
1-\bar{\lambda} & =(1-\lambda) \frac{\tr(F\rho_{in}^{b})}{\tr(F\rho_{in})}\ge0,
\end{align}
so $0\le\bar{\lambda}\le1$. Of course, if $\tr[\varphi(\rho_{in})]=1$
then $\bar{\lambda}=\lambda$ (compare \cite{GKM2006_OSID}).

From the above definition it follows that the frequently 
used argument that two equivalent
ensembles after nonlinear map lost this equivalence does not apply
to the map $\Phi$.
This means that $\Phi$ can be treated as the generalized form 
of the acceptable quantum operation.

\section{Admissible nonlinear evolutions}
\label{sec:nonlin_evolutions}

Now, the notion of convex quasi-linearity can be extended for 
deterministic time evolution. Namely, the transformation 
\begin{equation}
\rho(t) = \Phi_t [\rho_0], \qquad \rho_0=\rho(0),
\end{equation}
is a convex quasi-linear time evolution if $\Phi_t$ forms a semigroup, i.e.,
$\Phi_{t_1}\circ\Phi_{t_2}=\Phi_{t_1+t_2}$ and the condition
(\ref{convex_quasi-linear}) holds for all times $t$, i.e., for all 
$\rho_{0i}\in S$, $p_i\in\langle0,1\rangle$, 
$\sum_i p_i=1$ there exist such $p_i(t)$ that:
\begin{equation}
\Phi_t \big[\sum_i p_i \rho_{0i}\big] 
= \sum_i p_i(t) \Phi_t[\rho_{0i}]
\label{convex_quasi-linear_time}
\end{equation}
and $p_i(t)\in\langle0,1\rangle$, $\sum_i p_i(t)=1$.
In our paper \cite{RC2019_Nonlinearity_signaling}
we have found a model of a convex quasi-linear 
time evolution of a qubit. Here we show that there exists
a large class of natural evolutions fulfilling the above conditions.

\subsection{Quasi-linear time evolution generated by linear transformations}

According to the above discussion there are no formal objections to identify
time evolution of density operators with a family of time dependent extended
quantum operations.
Consequently we postulate the evolution of the density operator in the
form of the nonlinear map
\begin{equation}
\rho(t) = \Phi_t(\rho_0)  = 
\frac{\varphi_t(\rho_0)}{\tr[\varphi_t(\rho_0)]}
= \frac{\sum_{\alpha} K_\alpha(t) \rho_0 K_{\alpha}^{\dagger}(t)}{\tr[F(t)\rho_0]},
\label{nonlin_dyn_global}
\end{equation}
with
\begin{equation}
F(t) = \sum_{\alpha} K_{\alpha}^{\dagger}(t) K_{\alpha}(t).
\end{equation}
Here $\varphi_t(\rho_0)=\sum_{\alpha=0}^{\alpha_{max}} K_\alpha(t) 
\rho_0 K_{\alpha}^{\dagger}(t)$, $\tr(\rho_0)=1$. Notice that
if $F(t)=I$ then the evolution is linear.
The initial condition $\rho_0 = \rho(0)$ is related standardly with
$K_0(0)=I$ and $K_\alpha(0)=0$ for $\alpha=1,2,\dots,\alpha_{max}$.

Let us assume that the family of linear positive maps $\varphi_t$
satisfy $\varphi_\tau(\varphi_t(M))=\varphi_{\tau+t}(M)$ for each 
$M$, i.e., $\{\varphi_t\}$ forms a one parameter semigroup. 
Then using linearity of $\varphi_t$ and the definition of $\Phi_t$ we have
for each $\rho_0$
\begin{align}
\Phi_\tau(\Phi_t(\rho_0)) 
& = \frac{\varphi_\tau(\Phi_t(\rho_0))}{\tr[\varphi_\tau(\Phi_t(\rho_0))]}
\nonumber\\
& = \frac{\varphi_\tau\Big( \frac{\varphi_t(\rho_0)}{\tr[\varphi_t(\rho_0)]} \Big)}{\tr\Big[\varphi_\tau\Big( \frac{\varphi_t(\rho_0)}{\tr[\varphi_t(\rho_0)]} \Big) \Big]}\nonumber\\
& = \frac{\varphi_\tau(\varphi_t(\rho_0))}{\tr[\varphi_\tau(\varphi_t(\rho_0))]}
\nonumber\\
& =\frac{\varphi_{\tau+t}(\rho_0)}{\tr[\varphi_{\tau+t}(\rho_0)]}
\nonumber\\
& = \Phi_{\tau+t}(\rho_0),
\end{align}
i.e., $\Phi_\tau(\Phi_t(\rho_0))=\Phi_{\tau+t}(\rho_0)$.
We conclude that under our assumptions the family of
quantum operations $\Phi_t(\rho_0)$ forms a nonlinear realization of the
one-parameter semigroup realized in the convex set of the density matrices.
In particular, subfamily of trace-preserving evolutions 
(for $F(t)=I$) became linear.
Notice also that the above time evolution preserves convex quasi-linearity,
i.e.,
\begin{equation}
\Phi_t(\lambda \rho_{0}^{a} + (1-\lambda)\rho_{0}^{b})
= \bar{\lambda}(t) \Phi_t(\rho_{0}^{a}) 
+ (1-\bar{\lambda}(t)) \Phi_t(\rho_{0}^{b}),
\end{equation}
where $0\le\lambda\le 1$ and
\begin{align}
\bar{\lambda}(t) & =\lambda\frac{\tr[\varphi_t(\rho_{0}^{a})]}{\tr[\varphi_t(\rho_{0})]},\\
1-\bar{\lambda}(t) & =(1-\lambda) \frac{\tr[\varphi_t(\rho_{0}^{b})]}{\tr[\varphi_t(\rho_{0})]},
\end{align}
and $\bar{\lambda}(t)\le1$.

\subsection{Nonlinear extension of the
	 Gorini--Kossakowski--Sudarshan--Lindblad equation}

The linear dynamics of an open quantum system is described
by Gorini--Kossakowski--Sudarshan--Lindblad (GKSL) dynamical
semigroup which is generated by the GKSL generator.

Using  the infinitesimal form of the global time evolution as well as  the form of
the effect operator $F(t)$ and defining
$K_0(\delta t)\approx I+\delta t(G-iH)$,
where $G$ and $H$ are Hermitian while for
$\alpha=1,2,\dots,\alpha_{max}$, $K_\alpha(\delta t)\approx \sqrt{\delta t}L_\alpha$,
we obtain the generalization of the action of this generator to the form 
\begin{multline}
\dot{\rho} = \mathcal{L}_\Phi[\rho]
\equiv 
-i[H,\rho] + \{G,\rho\}
+ \sum_{\alpha=1} (L_\alpha \rho L_{\alpha}^{\dagger})\\
- \rho \tr\Big[ 
\rho \big( 2G + \sum_{\alpha=1} L_{\alpha}^{\dagger} L_\alpha  \big)
\Big]
\label{GKSL_nonlinear_general}
\end{multline}
(the detailed derivation of the above equation is given in Appendix).
Notice, that the nonlinearity of the dynamics of the density operator is rather weak:
It relies on the nonlinear coupling between $\rho$ (operator) and the $\rho$-dependent
trace $\tr\Big( 
\rho \big( 2G + \sum_{\alpha=1} L_{\alpha}^{\dagger} L_\alpha  \big)
\Big)$.

From the above dynamical equation it follows that infinitesimally
the operator $F(t)$ is generated by 
\begin{equation}
2G+\sum_{\alpha=1} L_{\alpha}^{\dagger} L_\alpha.
\end{equation}
We observe that for $F(t)=I$, i.e., for 
$2G+\sum_{\alpha=1} L_{\alpha}^{\dagger} L_\alpha=0$,
we recover the standard form of the GKSL generator:
\begin{equation}
\mathcal{L}[\rho] = -i[H,\rho]
+ \sum_{\alpha=1} \Big(
L_\alpha \rho L_{\alpha}^{\dagger}
-\frac{1}{2} \{ L_{\alpha}^{\dagger} L_\alpha,\rho \}
\Big).
\end{equation}
Notice, that  vanishing of the Lindblad generators (only $K_0$  is nonzero)
implies that the nonlinear Kraus evolution reduces to 
\begin{equation}
\rho(t) = 
\frac{K(t) \rho_{0} K(t)^\dagger}{\tr[K(t) \rho_{0} K(t)^\dagger]}
\label{rho_t}
\end{equation}
with
\begin{equation}
K(t) = e^{(G-iH)t},
\label{Kt_vanishing_Lindblad-1}
\end{equation}
where $G^\dagger=G$, $H^\dagger=H$
and we can restrict ourselves to the case of traceless generators $G$ and $H$.
Thus, the family of $K(t)$ operators forms an one-parameter subgroup of the
\textsf{SL}$(N,\mathbb{C})$ linear group.
The corresponding GKSL equation reduces to the nonlinear generalization of
the von Neumann equation, i.e., (compare \cite{SZ2013_non-Hermitian_q_dyn_two-level})
\begin{equation}
\dot{\rho} = -i[H,\rho] + \{G,\rho\}
-2 \rho \tr(G\rho)
\label{nonlinvN_rho}
\end{equation}
with the initial condition $\rho(0)=\rho_{0}$.
Let us observe that the pure states form for this evolution an invariant subset
in the convex set of density operators.
Indeed, we see that in this case
$\rho_{0}=\ket{\psi}\bra{\psi}$ with $\bracket{\psi}{\psi}=1$,
so the evolution equation takes the form 
\begin{equation}
\rho(t) = \frac{K(t)\ket{\psi}\bra{\psi}K(t)^\dagger}{\bra{\psi}K(t)^\dagger K(t)\ket{\psi}}.
\end{equation}
Therefore
\begin{equation}
\rho(t)^2 = \Bigg(\frac{K(t)\ket{\psi}\bra{\psi}K(t)^\dagger}{\bra{\psi}K(t)^\dagger K(t)\ket{\psi}}\Bigg)^2=\rho(t),
\end{equation}
i.e., $\rho(t)$ is a pure state.
Therefore, we can find a counterpart of Eq.~(\ref{nonlinvN_rho}) for
state vectors. However, the corresponding equation describing the evolution
of a state vector is not uniquely determined by Eq.~(\ref{nonlinvN_rho}).
Indeed, the whole family of equations of the form
\begin{equation}
\frac{d}{dt}\ket{\psi} = 
\Big( -iH + G - \frac{\bra{\psi}G\ket{\psi}}{\bracket{\psi}{\psi}}I
+ i  \kappa I \Big)\ket{\psi},
\label{nonlinvN_psi}
\end{equation}
where $\kappa$ is an arbitrary real function of $\ket{\psi}$,
leads to the evolution equation (\ref{nonlinvN_rho}) for density matrices.
It is worth to mention here that in general the evolution equation for a state vector 
also does not determine uniquely the evolution equation for density matrices.

Evolution of the form (\ref{nonlinvN_psi}) for pure states with $\kappa=0$
and $G=kH$ was discussed by Gisin in \cite{Gisin1981_dissipative_q_dyn}
and subsequently with general $G$ and $\kappa=0$ in
\cite{Gisin1983_Irreversible_q_dyn}. 
Evolution similar to (\ref{nonlinvN_rho}) was also considered in a specific
experimental context of interaction of a two-level atom with maser photons
\cite{BESW1994}.
Eq.~(\ref{nonlinvN_rho}) was also proposed in
\cite{BG2012_non-Hermitian-gain-loss} as an evolution equation in presence of gain and loss while
in \cite{KAU2017_Non-Hermitian-information-retrieval} the information retrieval
during such an evolution was discussed.

The fact that equation (\ref{nonlinvN_rho}) preserves the coherence of pure 
states means that 
in the density matrix evolution
this part of the master equation (\ref{GKSL_nonlinear_general})
plays different role than Lindblad operators. 
We will discuss this question later.

The generalized GKLS master equation (\ref{GKSL_nonlinear_general}) admits 
also evolutions with time-dependent generators; 
in particular in (\ref{nonlinvN_rho}) and (\ref{nonlinvN_psi}) $H$ and $G$ 
can be time-dependent. 
In such a case the global solutions have different form than those with 
time-independent generators. 
Such evolutions are very interesting because they can contain structural instability points 
\cite{AAIS1991}.

\section{Nonlinear qubit evolution}
\label{sec:qubit}

In this section we discuss qubit evolution in our model.
Under the special choice of the von Neumann nonlinear equation it 
was also investigated recently in \cite{SZ2013_non-Hermitian_q_dyn_two-level,SZ2015_non-Hermitian_time_correl,Zloshchastiev2015_non-Hermitian_pure_states,GCKM_su11_Hamiltonian}.
In this case the initial density matrix can be taken in the following form:
\begin{equation}
\rho(0) = \frac{1}{2} (I+\vecg{\xi}\cdot\vecg{\sigma}),
\label{initial_qubit}
\end{equation}
where $\vecg{\xi}$ is a real vector satisfying the condition
$\vecg{\xi}^2\le1$ and $\vecg{\sigma}$ is the triple of the Pauli matrices.

\subsection{The case of vanishing Lindblad generators ($L_\alpha=0$)}
\label{subsec:Lzero}

The \textsf{SL}$(2,\mathbb{C})$ nonlinear evolution of $\rho$ in the case of vanishing of the Lindblad  generators has the form
\begin{equation}
\rho(t) = \frac{1}{2} (I+\vec{n}(t)\cdot\vecg{\sigma})
 = \frac{K(t)\rho(0) K(t)^\dagger}{\tr[K(t)\rho(0) K(t)^\dagger]},
 \label{nt_no_Lindblads_gen}
\end{equation}
where the evolution matrix reads
\begin{equation}
K(t) = e^{(G-iH)t},
\label{Kt_no_Lindblads_gen}
\end{equation}
with $G=\tfrac{1}{2}\vec{g}\cdot\vecg{\sigma}$,
$H=\tfrac{1}{2}\vecg{\omega}\cdot\vecg{\sigma}$, and $\vec{g}$, $\vecg{\omega}$
are fixed real vectors. Moreover we assume $\vec{n}(0)=\vecg{\xi}$.

This evolution can be represented on the Bloch ball as a nonlinear
realization of one-parameter subgroup of the orthochronous Lorentz group
homomorphic to \textsf{SL}$(2,\mathbb{C})$.
Notice that the quantities $C_1=\vec{g}\cdot\vecg{\omega}$ and 
$C_2=\vec{g}^2-\vecg{\omega}^2$ are invariant under inner automorphisms
of the \textsf{SL}$(2,\mathbb{C})$. Therefore, values of $C_1$ and $C_2$
determine different types of evolution.

We get explicitly 
\begin{equation}
K(t) = e^{(G-iH)t} = a I + b (\vecg{\alpha}\cdot\vecg{\sigma}),
\end{equation}
where $\vecg{\alpha}= \vec{g}-i\vecg{\omega}$ and 
for $\vecg{\alpha}^2\not=0$:
\begin{equation}
a=\cosh(\tfrac{t}{2}\sqrt{\vecg{\alpha}^2}),\quad 
b=\tfrac{1}{\sqrt{\vecg{\alpha}^2}}\sinh(\tfrac{t}{2}\sqrt{\vecg{\alpha}^2}),
\label{ab_alpha_not_0}
\end{equation}
while for $\vecg{\alpha}^2=0$:
\begin{equation}
a=1,\quad b=\tfrac{t}{2}.
\label{ab_alpha_0}
\end{equation}
The general form of $\vec{n}(t)$ can be found from 
Eq.~(\ref{nt_no_Lindblads_gen}). 
After simple but rather lengthy calculation we obtain:
\begin{widetext}
\begin{multline}
	\vec{n}(t) = \Big\{ aa^* + bb^*\big(g^2+\omega^2 - 
	2 (\vecg{\omega}\times\vec{g})\cdot\vecg{\xi}\big)
	+(ab^*+a^*b)(\vec{g}\cdot\vecg{\xi})
	+i(ab^*-a^*b)(\vecg{\omega}\cdot\vecg{\xi})\Big\}^{-1}\times
	\Big\{  [aa^*-bb^*(g^2+\omega^2)]\vecg{\xi}\\
	+ [ab^*+a^*b+2bb^*(\vec{g}\cdot\vecg{\xi})]\vec{g} 
	+[i(ab^*-a^*b)+2bb^*(\vecg{\omega}\cdot\vecg{\xi})]\vecg{\omega}
	-2bb^*(\vec{g}\times\vecg{\omega})
	-i(ab^*-a^*b)(\vec{g}\times\vecg{\xi})\\
	+(ab^*+a^*b)(\vecg{\omega}\times\vecg{\xi})
	\Big\}.
	\label{nt-general-a-b}
\end{multline}
If we restrict our attention to the case when
$C_1=\vecg{\omega}\cdot\vec{g}=0$
then $\vecg{\alpha}^2\in{\mathbb{R}}$ and 
for the sake of brevity we introduce the following notation:
\begin{equation}
	\Omega = \sqrt{|g^2-\omega^2|}.
	\label{Omega_def}
\end{equation}
The considered case divides into three sub-classes:\\
(i) $g^2=\omega^2$ ($\vecg{\alpha}^2=0$), $a,b$ are given 
in Eq.~(\ref{ab_alpha_0})  and 
\begin{equation}
	\vec{n}(t) = \frac{(1-\tfrac{1}{2}\omega^2 t^2)\vecg{\xi} 
		+ [t+\tfrac{1}{2} (\vec{g}\cdot\vecg{\xi}) t^2]\vec{g} 
		+\tfrac{1}{2}(\vecg{\omega}\cdot\vecg{\xi}) t^2 \vecg{\omega}
		+\tfrac{1}{2} t^2 (\vecg{\omega}\times\vec{g})
		+t(\vecg{\omega}\times\vecg{\xi})}{1+t(\vec{g}\cdot\vecg{\xi})
		+\tfrac{1}{2} t^2 (\omega^2 - (\vecg{\omega}\times\vec{g})\cdot\vecg{\xi})},
	\label{Lzero_gro}
\end{equation}
(ii) $g^2>\omega^2$, $\sqrt{\vecg{\alpha}^2}=\Omega$ 
(compare Eq.~(\ref{Omega_def})), 
$a=\cosh(\tfrac{1}{2}\Omega t)$,
$b=\tfrac{1}{\Omega}\sinh(\tfrac{1}{2}\Omega t)$ and
\begin{equation}
	\vec{n}(t) =  \frac{\big( g^2 - \omega^2 \cosh(\Omega t) \big) \vecg{\xi}
		+ \Omega \sinh(\Omega t) (\vec{g} + \vecg{\omega}\times \vecg{\xi})
		- \big( 1- \cosh(\Omega t) \big) 
		[(\vec{g}\cdot\vecg{\xi})\vec{g}
		+ (\vecg{\omega}\cdot\vecg{\xi})\vecg{\omega}
		+ \vecg{\omega}\times\vec{g}]}{g^2 \cosh(\Omega t) - \omega^2
		+ \big(1- \cosh(\Omega t) \big) (\vecg{\omega}\times\vec{g})\cdot\vecg{\xi} 
		+ \Omega \sinh(\Omega t) (\vec{g}\cdot\vecg{\xi}) },
	\label{Lzero_gwo}
\end{equation}
(iii) $\omega^2>g^2$, $\sqrt{\vecg{\alpha}^2}=i \Omega$ 
(compare Eq.~(\ref{Omega_def})), 
$a=\cos(\tfrac{1}{2}\Omega t)$,
$b=\tfrac{1}{\Omega}\sin(\tfrac{1}{2}\Omega t)$ and
\begin{equation}
	\vec{n}(t) = \frac{\big( \omega^2 \cos(\Omega t) - g^2 \big)\vecg{\xi}
		+ \Omega \sin(\Omega t) (\vec{g}+\vecg{\omega}\times\vecg{\xi})
		+ \big( 1- \cos(\Omega t)\big) 
		[(\vec{g}\cdot\vecg{\xi})\vec{g}
		+ (\vecg{\omega}\cdot\vecg{\xi})\vecg{\omega}
		+ \vecg{\omega}\times\vec{g}]}{\omega^2 - g^2 \cos(\Omega t)
		- \big( 1-\cos(\Omega t)\big) (\vecg{\omega}\times\vec{g})\cdot\vecg{\xi}
		+ \Omega \sin(\Omega t) (\vec{g}\cdot \vecg{\xi})}.
	\label{Lzero_gmo}
\end{equation}
\end{widetext}

To simplify our example, in what follow we put 
$\vecg{\omega}=(0,0,\omega)$ and $\vec{g}=(g,0,0)$, $\omega,g>0$.
In this case the generator $H+iG$ satisfies the $PT$-symmetry condition and was 
used in \cite{BG2012_non-Hermitian-gain-loss}.

Let us calculate the time evolution of the probability 
of finding the evolved state $\rho(t)$ in eigenvectors
of the Hamiltonian $H$. 
For $\vecg{\omega}=(0,0,\omega)$
the normalized eigenvectors of the Hamiltonian $H=\tfrac{1}{2}\sigma_3$ 
have the form:
\begin{equation}
a(0)_+ = \begin{pmatrix}
1 \\ 0
\end{pmatrix},\quad
a(0)_- = \begin{pmatrix}
0 \\ 1
\end{pmatrix},
\end{equation}
thus the corresponding projectors correspond to the
$\vecg{\xi}_\pm = (0,0,\pm1)$. We obtain:\\
$g=\omega$
\begin{align}
p_+(t) & = \frac{1+n_3(t)}{2} =
\frac{4+(gt)^2}{4+2(gt)^2},\\
p_-(t) & = \frac{1-n_3(t)}{2} =
\frac{(gt)^2}{4+2(gt)^2}.
\end{align}
In the asymptotic limit $p_\pm(\infty)=\frac{1}{2}$.\\
$g>\omega$
\begin{align}
p_+(t) & = \frac{1+n_3(t)}{2} =
\frac{1}{2} \frac{g^2\cosh(\sqp t)+g^2-2\omega^2}{g^2\cosh(\sqp t)-\omega^2},\\
p_-(t) & = \frac{1-n_3(t)}{2} =
\frac{1}{2} \frac{g^2\cosh(\sqp t)-g^2}{g^2\cosh(\sqp t)-\omega^2}.
\end{align}
In the asymptotic limit $p_\pm(\infty)=\frac{1}{2}$.\\
$g<\omega$
\begin{align}
p_+(t) & = \frac{1+n_3(t)}{2} =
\frac{1}{2} \frac{2\omega^2-g^2\cos(\sqm t)-g^2}{\omega^2-g^2\cos(\sqm t)},
\label{prob-gmo-p}\\
p_-(t) & = \frac{1-n_3(t)}{2} =
\frac{1}{2} \frac{g^2-g^2\cos(\sqm t)}{\omega^2-g^2\cos(\sqm t)}.
\label{prob-gmo-m}
\end{align}

The case $g<\omega$ (when $\sqm=\sqrt{\omega^2 - g^2}$)
is the most interesting one (compare with \cite{GCKM_su11_Hamiltonian}). 
In this case the probability of finding the evolved state in one of the eigenstates 
of the Hamiltonian $H$, $a_-=(0,1)$, is given by (\ref{prob-gmo-m}).
For comparison, let us remind that the counterpart of the probability
(\ref{prob-gmo-m}) obtained for the standard Rabi oscillations
(i.e. in the case of evolution governed by the Hamiltonian
$H=\tfrac{\omega}{2}\sigma_3 + \tfrac{g}{2}\sigma_1$)
has the form \cite{Cohen1991_book_QM}
\begin{equation}
p_{-}^{\mathsf{Rabi}} = \frac{g^2}{2(g^2+\omega^2)}
\big(1-\cos( t\sqrt{g^2+\omega^2}) \big).
\label{pmRabi}
\end{equation}
It is interesting to notice that both probabilities, $p_-(t)$ (\ref{prob-gmo-m})
and $p_{-}^{\mathsf{Rabi}}$ (\ref{pmRabi}), for fixed $g$ and $\omega$ 
attain the same maximal value equal to
\begin{equation}
p_{-max} = \frac{g^2}{g^2 + \omega^2}.
\end{equation}

\begin{figure}
\includegraphics[width=0.95\columnwidth]{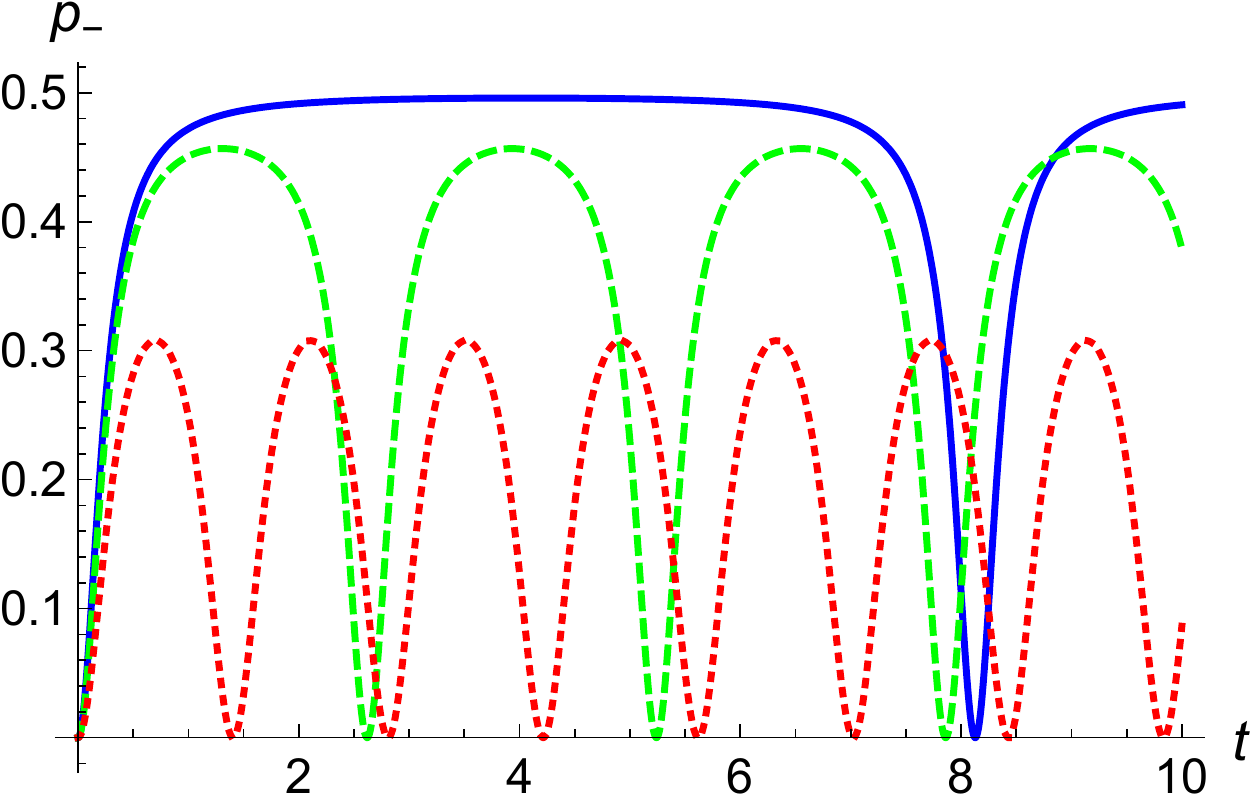}
\caption{The probability (\ref{prob-gmo-m}) for $\omega=6$
	 and $g=4$ (red, dotted line),
$g=5.5$ (green, dashed line) and $g=5.95$ (blue, solid line).}
\label{fig1}
\end{figure}

\begin{figure}
\includegraphics[width=0.95\columnwidth]{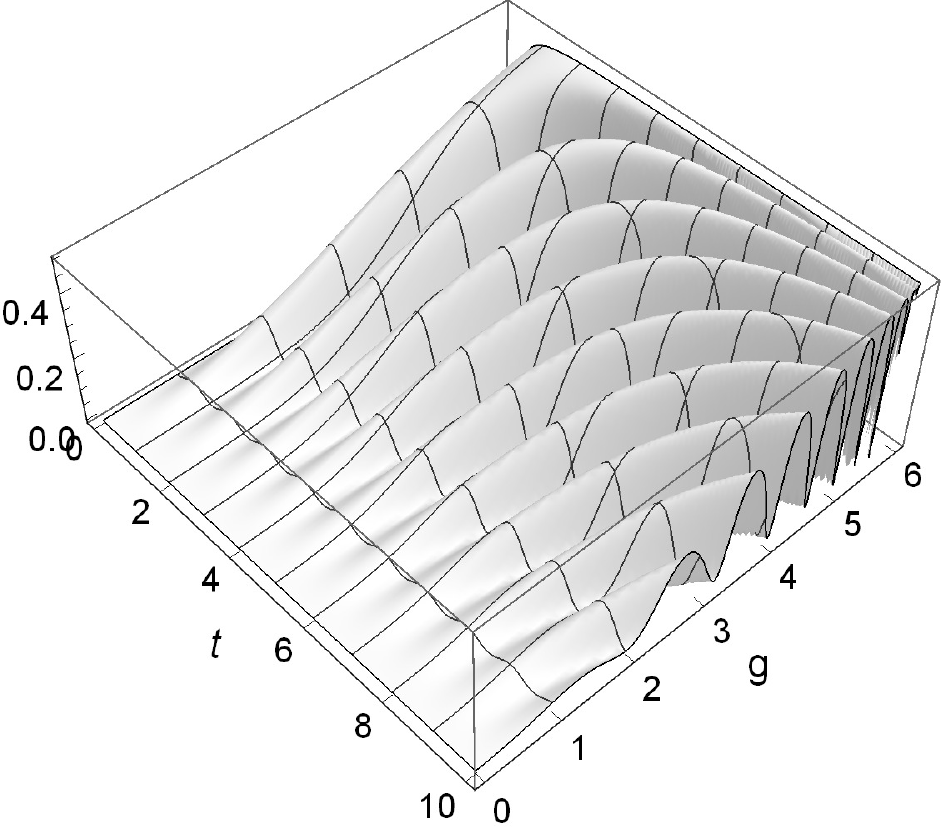}
\caption{The probability (\ref{prob-gmo-m}) for $\omega=6$.}
\label{fig2}
\end{figure}

\begin{figure}
\includegraphics[width=0.95\columnwidth]{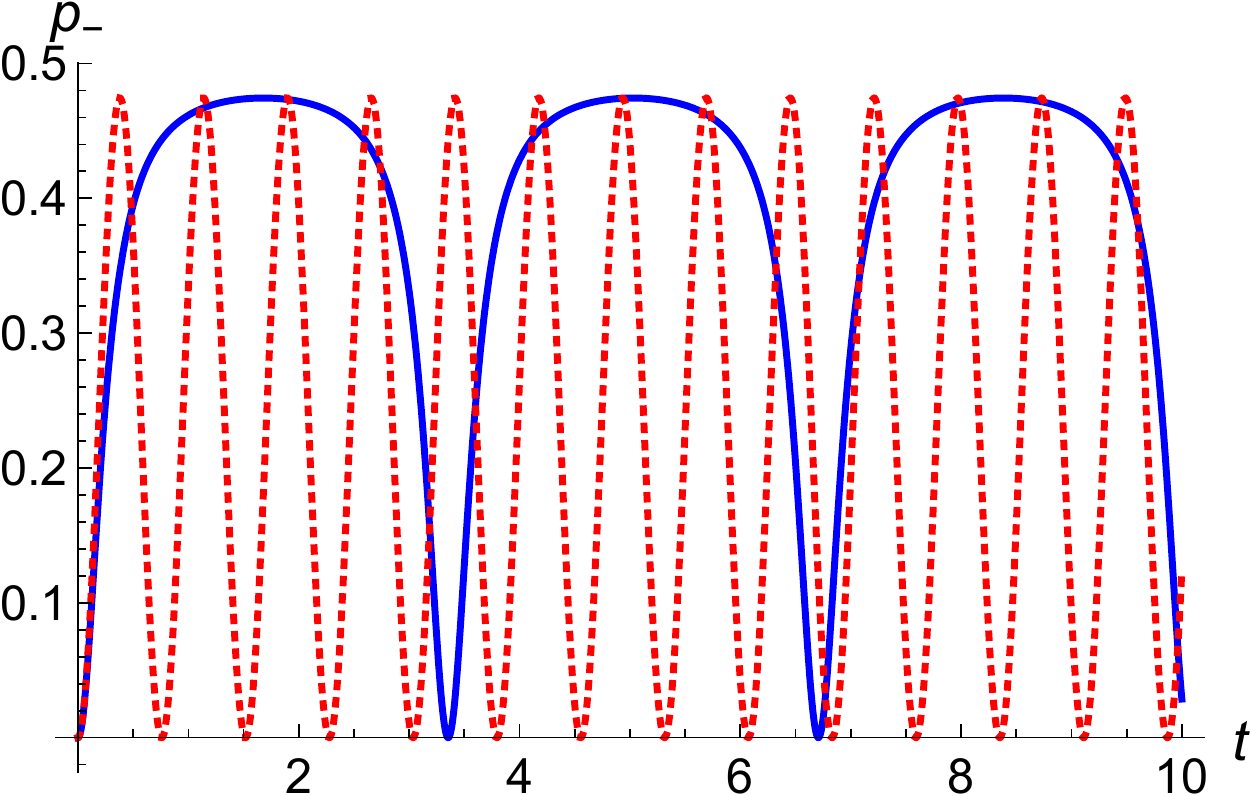}
\caption{Comparison of the probability (\ref{prob-gmo-m})
	(blue, solid line) with 
	the probability obtained it the case of standard Rabi oscillations (\ref{pmRabi})
	(red, dotted line). Both curves are drown for $\omega=6$ and $g=5.7$.}
\label{fig3}
\end{figure}

Below we illustrate the behavior of the probabilities
(\ref{prob-gmo-m})
and compare it with the probabilities obtained for the standard Rabi 
oscillations (\ref{pmRabi}).
In Figs.~\ref{fig1} and \ref{fig2} we present the probability
(\ref{prob-gmo-m}) for $g<\omega$. 
In Fig.~\ref{fig3} we compare the probability (\ref{prob-gmo-m}) with 
the probability obtained it the case of standard Rabi oscillations (\ref{pmRabi})
(for the same values of parameters $\omega$ and $g$).

\begin{figure}
	\includegraphics[width=0.95\columnwidth]{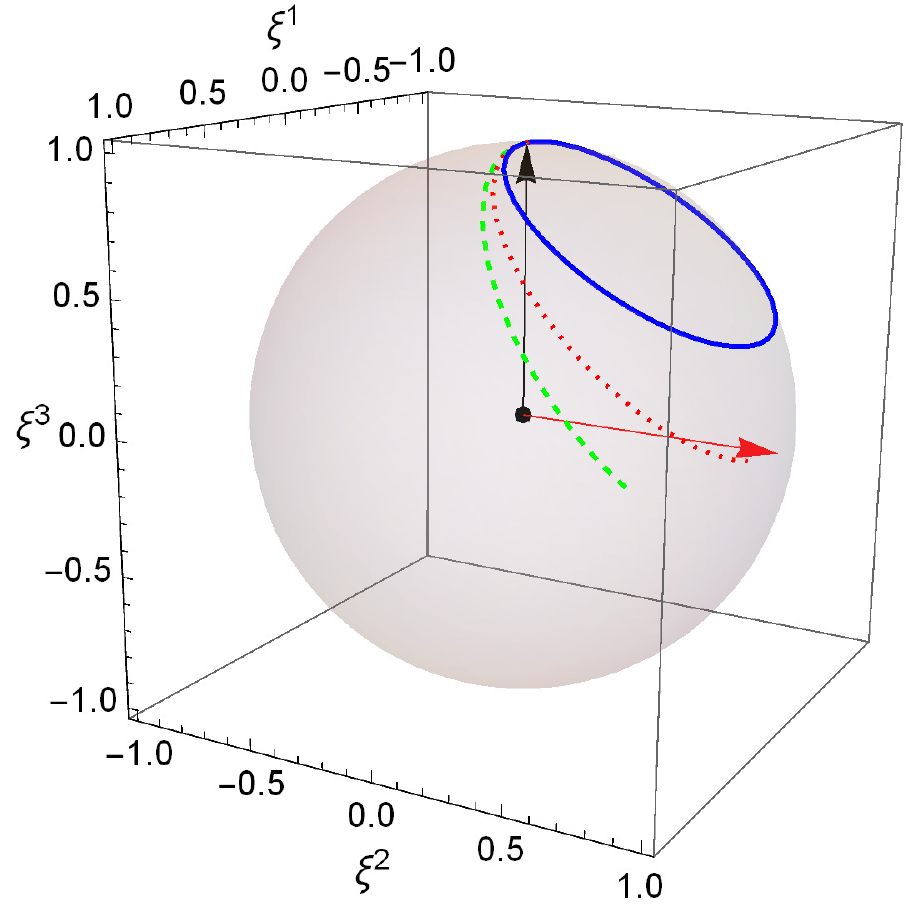}
	\caption{The trajectory of the Bloch vector under the evolution
		(\ref{nt_no_Lindblads_gen}) with the initial condition 
		$\vecg{\xi}=(0,0,1)$. 
		Blue (solid) line corresponds to the case $\omega>g$ (for the plot we 
		set $\omega=6$, $g=4$).
		Red (dotted) line corresponds to the case $\omega=g$ (for the plot we 
		set $\omega=6$).
		Green (dashed) line corresponds to the case $\omega<g$ (for the plot we 
		set $\omega=6$, $g=8$).
		Black (vertical) arrow represents $\vecg{\xi}$, red arrow represents the vector $(0,1,0)$.}
	\label{fig4}
\end{figure}

In Fig.~\ref{fig4} we show the trajectories of
the Bloch vector under the evolution
(\ref{nt_no_Lindblads_gen}) with the initial condition 
$\vecg{\xi}=(0,0,1)$ 
and with
$\vecg{\omega}=(0,0,\omega)$ and $\vec{g}=(g,0,0)$, $\omega,g>0$
for all three cases $g>\omega$ [Eq.~(\ref{Lzero_gwo})], $g=\omega$ [Eq.~(\ref{Lzero_gro})], and
$g<\omega$ [Eq.~(\ref{Lzero_gmo})]. 
Since the initial state is pure $|\vecg{\xi}|=1$,
the vector $\vec{n}(t)$ has length 1 for all $t$. It means that the curves
in Fig.~\ref{fig4} are situated on the Bloch sphere.
For comparison, in Fig.~\ref{fig6} we present similar trajectories for the initial
state $\rho(0)=\tfrac{1}{2}I$ ($\vecg{\xi}=(0,0,0)$ and again
$\vecg{\omega}=(0,0,\omega)$ and $\vec{g}=(g,0,0)$, $\omega,g>0$)
[for calculations we used formulas
(\ref{Lzero_gro},\ref{Lzero_gwo},\ref{Lzero_gmo})]. 
In this case the curves are inside the Bloch sphere on the plane
$\xi^3=0$.

\begin{figure}
	\includegraphics[width=0.95\columnwidth]{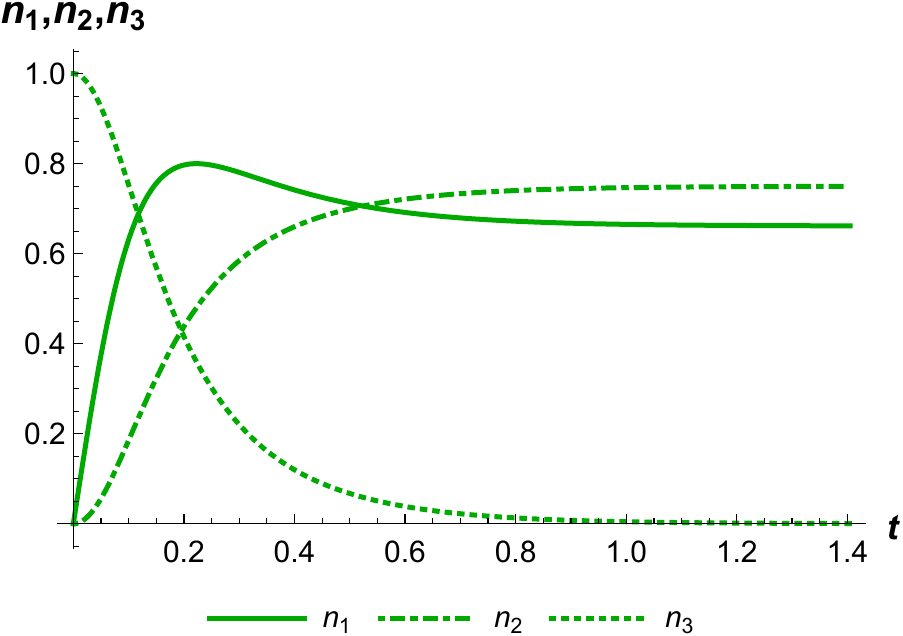}
	\caption{The behavior of the components of the Bloch vector 
		$n_1(t)$, $n_2(t)$,	$n_3(t)$ under the evolution  
		depicted with green, dashed line in	Fig.~\ref{fig4}
		[evolution (\ref{nt_no_Lindblads_gen}) with the initial condition 
		$\vecg{\xi}=(0,0,1)$
		in the case $\omega<g$ (for the plot we 
		set $\omega=6$, $g=8$)].}
	\label{fig5}
\end{figure}

\begin{figure}
	\includegraphics[width=0.95\columnwidth]{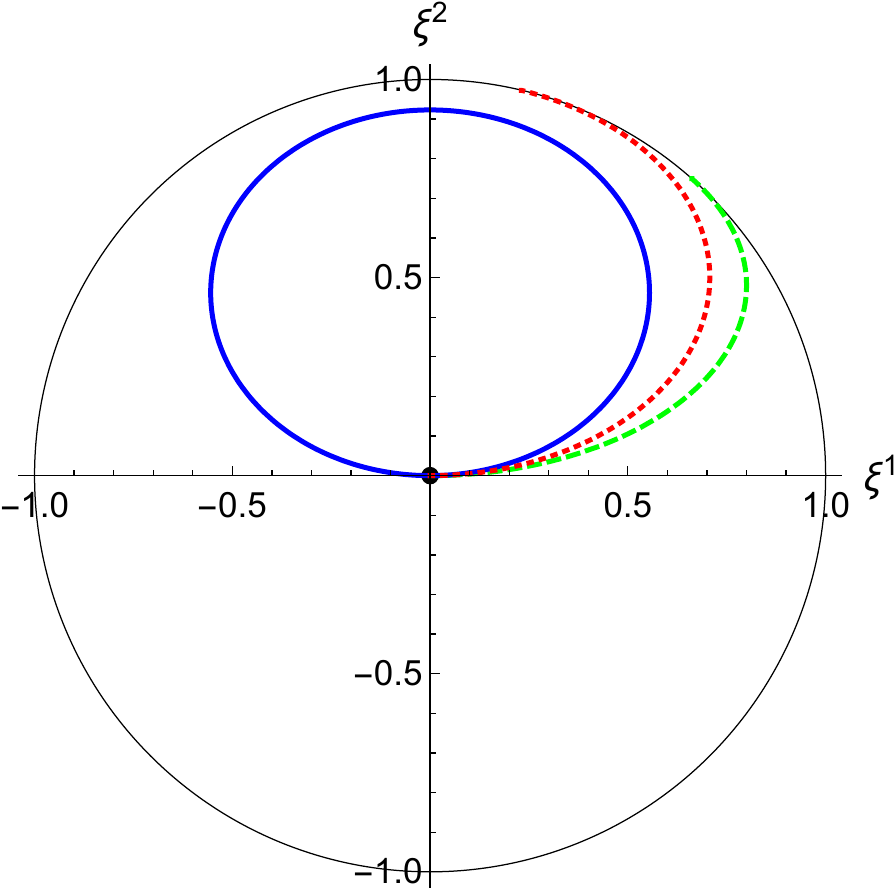}
	\caption{The trajectory of the Bloch vector under the evolution
		(\ref{nt_no_Lindblads_gen}) with the initial condition 
		$\vecg{\xi}=(0,0,0)$, i.e., for the initial state $\rho(0)=\tfrac{1}{2}I$.
		Trajectories start at the point $(0,0,0)$ and are situated in the plane
		$\xi^3=0$. Therefore, we present the section of the Bloch
		sphere with this plane. 
		Blue (solid) line corresponds to the case $\omega>g$ (for the plot we 
		set $\omega=6$, $g=4$).
		Red (dotted) line corresponds to the case $\omega=g$ (for the plot we 
		set $\omega=6$).
		Green (dashed) line corresponds to the case $\omega<g$ (for the plot we 
		set $\omega=6$, $g=8$).}
	\label{fig6}
\end{figure}

\paragraph{Asymptotes}

In this subsection we discuss asymptotic behavior of quasilinear evolutions using 
as a model the explicit global solution (\ref{nt-general-a-b}). 
Analyzing time dependence of the Bloch vector $\vec{n}(t)$ 
we obtain the following classes of asymptotes:
\begin{widetext}
(i) $\vecg{\omega}\cdot \vec{g}\not=0$
\begin{equation}
\vec{n}(\infty) = 
\frac{[x^2+y^2-(g^2+\omega^2)]\vecg{\xi}
+2(x+\vec{g}\cdot\vecg{\xi})\vec{g}
+2(-y+\vecg{\omega}\cdot\vecg{\xi})\vecg{\omega}
+2[\vecg{\omega}\times\vec{g} + y(\vec{g}\times\vecg{\xi}) + x(\vecg{\omega}\times\vecg{\xi})]}{[x^2+y^2+(g^2+\omega^2)]
-2[(\vecg{\omega}\times\vec{g})\cdot\vecg{\xi}
-x(\vec{g}\cdot\vecg{\xi})
+y(\vecg{\omega}\cdot\vecg{\xi})]},
\end{equation}
where
\begin{equation}
x=\tfrac{1}{\sqrt{2}} \sqrt{g^2-\omega^2 + 
	\sqrt{(g^2-\omega^2)^2 + (2g\omega\cos\vartheta)^2}},
\quad
y=\tfrac{-\mathrm{sign}\,(g\omega\cos\vartheta)}{\sqrt{2}} \sqrt{\omega^2 - g^2+ 
	\sqrt{(g^2-\omega^2)^2 + (2g\omega\cos\vartheta)^2}},
\end{equation}
so
\begin{equation}
x^2+y^2 = \sqrt{(g^2-\omega^2)^2 + (2g\omega\cos\vartheta)^2}
\end{equation}
and $\vartheta$ is the angle between $\vecg{\omega}$ and $\vec{g}$.\\
(ii) $\vecg{\omega}\cdot\vec{g} = 0$ and $g>\omega$
\begin{equation}
\vec{n}(\infty) = \frac{-\omega^2 \vecg{\xi} 
+ [\sqrt{g^2-\omega^2} + (\vec{g}\cdot\vecg{\xi})]\vec{g}
+ (\vecg{\omega\cdot\vecg{\xi}})\vecg{\omega}
+ (\vecg{\omega}\times\vec{g})
+ \sqrt{g^2-\omega^2}(\vecg{\omega}\times\vecg{\xi})}{g^2 
- (\vecg{\omega}\times\vec{g})\cdot\vecg{\xi}
+ \sqrt{g^2-\omega^2}(\vec{g}\cdot\vecg{\xi})}.
\label{n_asymptotic_ii}
\end{equation}
(iii) $\vecg{\omega}\cdot\vec{g} = 0$ and $g=\omega$
\begin{equation}
\vec{n}(\infty) = \frac{2(\vecg{\omega}\times\vec{g})
- (g^2+\omega^2)\vecg{\xi}
+ 2(\vec{g}\cdot\vecg{\xi})\vec{g}
+ 2(\vecg{\omega}\cdot\vecg{\xi})\vecg{\omega}}{(g^2+\omega^2)
-2(\vecg{\omega}\times\vec{g})\cdot\vecg{\xi}}.
\label{n_asymptotic_iii}
\end{equation}
(iv) In the case $\vecg{\omega}\cdot\vec{g} = 0$ and $g<\omega$
the evolution is oscillatory and there are no asymptotes.
\end{widetext}
It is worth to stress that almost all quasilinear evolutions (\ref{nt-general-a-b}) 
of qubit attain stationary states. 
This mean that the operator  $G=\tfrac{1}{2} \vec{g}\cdot\vecg{\sigma}$ damps
oscillations generated by the Hamiltonian
$H=\tfrac{1}{2}\vecg{\omega}\cdot\vecg{\sigma}$. 
The only exception is the case when $\vecg{\omega}$ is perpendicular to $\vec{g}$ 
under condition $\omega>g$. 
Notice that this damping does not increase entropy of pure initial states 
($\vec{n}^2=1$ is the evolution invariant). 
Notice that in the neighborhood of the point given 
by the conditions $\vecg{\omega}\cdot\vec{g}=0$  and $\omega=g$
the character of the evolution changes drastically. It means that this point
can be  identified with the so called point of structural instability
\cite{AAIS1991}.
This point is especially important in analysis of evolutions with time dependent 
$\vecg{\omega}$ and/or $\vec{g}$ (see Sec.~\ref{sec:applications}).
A physical meaning of the vectors $\vecg{\omega}$ and $\vec{g}$ depends on 
the physical context (see, e.g., Sec.~\ref{sec:applications}).

It can be shown that in the case (i) and for $\omega\gg g$ the asymptotic state is 
very close to the Bloch vector determining one of eigenstates of the Hamiltonian $H$,
i.e., the state of the system approaches to a stationary state. 
In particular, for $\vecg{\omega}$ and $\vec{g}$ parallel or anti-parallel and 
$\omega>g$ the asymptotic state is exactly the stationary state
and it does not depend on the initial state $\vecg{\xi}$.

\subsection{The case of time-dependent $\vec{g}$}
\label{sec:time-dependent-g}

\begin{figure*}
	\subfigure[]{%
		\label{7a}
		\includegraphics[width=.45\textwidth]{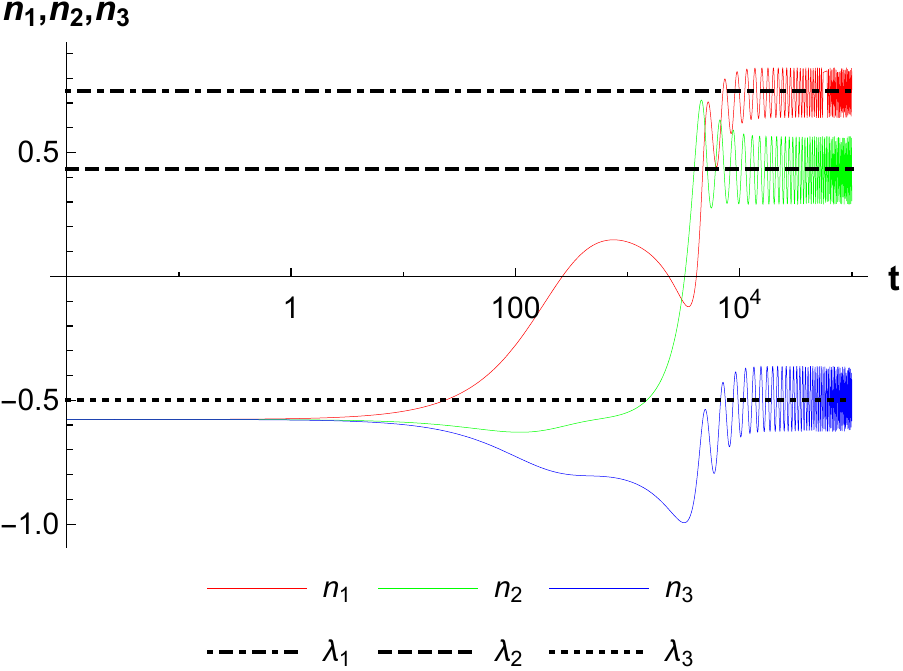}}
	\hspace{0.03\textwidth}
	\subfigure[]{%
		\label{7b}
		\includegraphics[width=.45\textwidth]{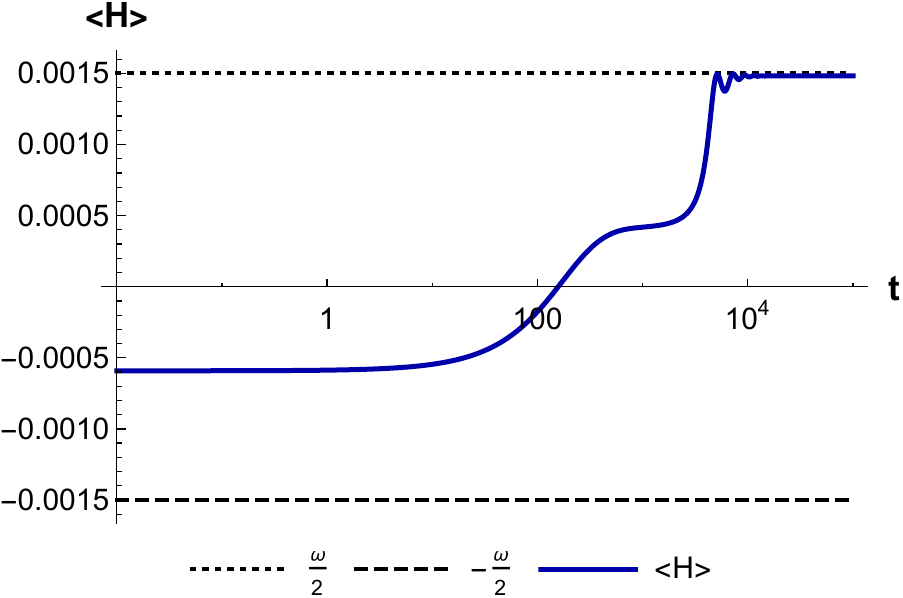}}
	\caption{The evolution of components of the Bloch vector (Fig.~\ref{7a})
	and the average energy $\langle H \rangle$ (Fig.~\ref{7b})
	for time-dependent $g$ (example discussed in Sec.~\ref{sec:time-dependent-g}).
	$t$-axis in logarithmic scale.}
	\label{fig7}
\end{figure*}

Let us consider an example of a qubit evolution governed by the nonlinear 
von Neumann equation (\ref{nonlinvN_psi}) in the case of the time-dependence 
of the vector $\vec{g}$. 
Let the vectors $\vecg{\omega}$ and $\vec{g}$ are oriented in the polar coordinates 
by the angles ($\theta =2\pi/3$, $\phi = \pi/6$) and ($\eta=4 \pi/3$, $\varepsilon =2 \pi/3$),
respectively, so the angle between $\vecg{\omega}$ and $\vec{g}$ is about 
of $75^\circ$.
We choose $\omega=0.003$,
$g=q[1-(1-e^{-\nu t})^2]$ (inverted Morse potential),  $q=0.007$ and $\nu=0.0005$. 
In that case Bloch vectors of stationary states of
the Hamiltonian have the form 
$\vecg{\lambda}_\pm = \pm(0.75,0.433,-0.5)$. 
Furthermore, the time coordinate $t_{in}$ of the structural instability point 
is determined by the condition $\omega=g(t_{in})$ so its value is $t_{in}=2821$
in this case.
In Fig.~\ref{7a} we presented the evolution of a qubit for the above parameters 
under initial condition $\vec{n}(0)=(-\tfrac{1}{\sqrt{3}},-\tfrac{1}{\sqrt{3}},-\tfrac{1}{\sqrt{3}})$.
We observe a strong damping of oscillations before the instability point and 
a rapid conversion to an weakly oscillatory state after this time. 
As we see, the averages of the components of the Bloch vector $\vec{n}$ 
for this outgoing state are very closed to components of
$\vecg{\lambda}_+$, represented by 
the dashed, dotted, dashed-dotted lines because 
for $t\gg t_{in}$ $\omega=g(t_{in})\gg g(t)$.
Next, in Fig.~\ref{7b} we present time-dependence of the average energy. 
We see that the energy flow is in this case directed to the qubit system. 
However, if we invert the direction of the vector $\vec{g}$ the average 
energy at the beginning goes up but next goes down to the lowest value
$-\tfrac{\omega}{2}$.
Thus we cannot identify the interaction of the system with the environment as dissipative.

\subsection{The case with nonzero Lindblad operators}

\begin{figure*}
	\subfigure[]{%
		\label{8a}
		\includegraphics[width=.45\textwidth]{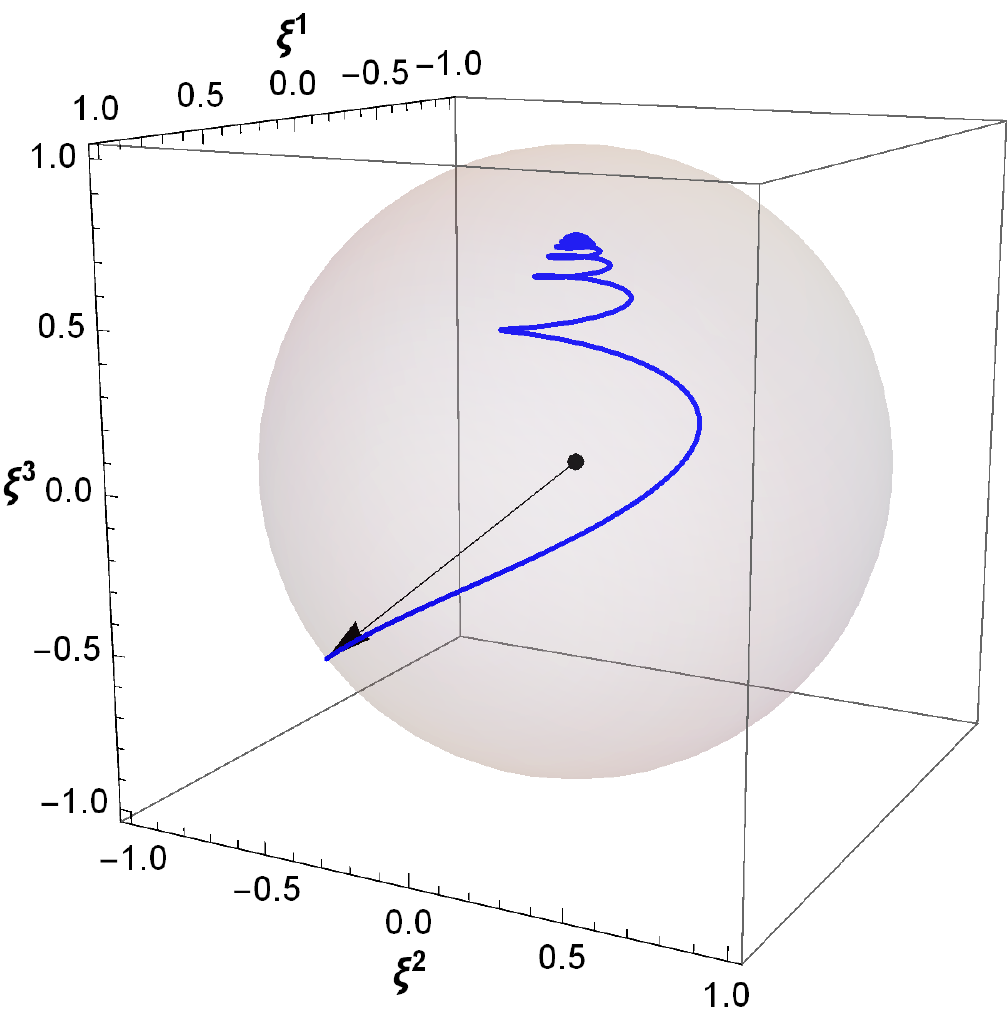}}
	\hspace{.03\textwidth}
	\subfigure[]{%
		\label{8b}
		\includegraphics[width=.45\textwidth]{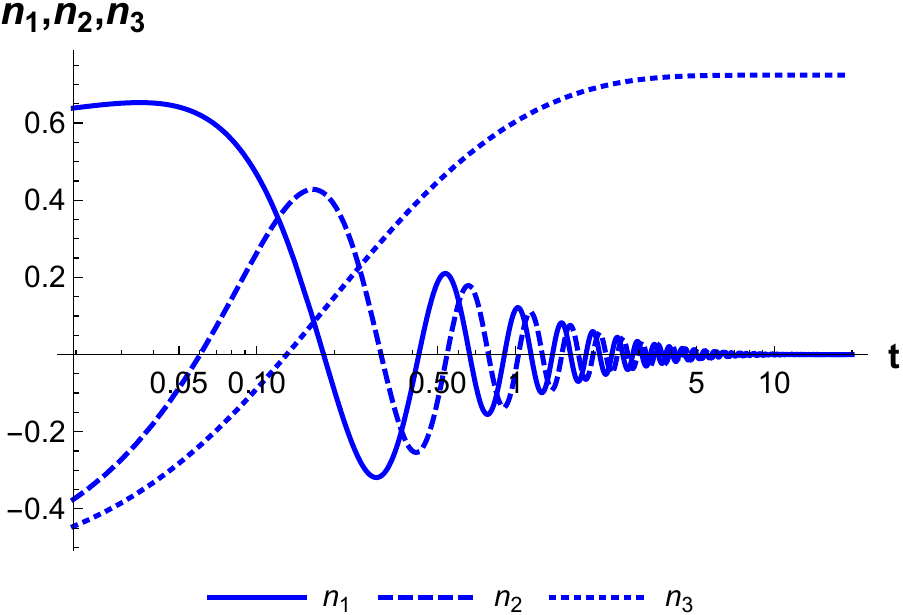}}
	\caption{The trajectory of the Bloch vector (Fig.~\ref{8a})
		and the values of the components of the Bloch vector 
		$n_1(t)$ (solid line), 
		$n_2(t)$ (dashed line),
		$n_3(t)$ (dotted line) (Fig.~\ref{8b})
		under the evolution
		(\ref{n_evol_nonzeroL}) with the initial condition 
		$\vecg{\xi}=(1/\sqrt{3},-1/\sqrt{3},-1/\sqrt{3})$.
		We assume that $g=-0.5$, $\omega=13$, $l=2.5$.
		$t$-axis in logarithmic scale.}
	\label{fig8}
\end{figure*}

\begin{figure}
	\includegraphics[width=0.95\columnwidth]{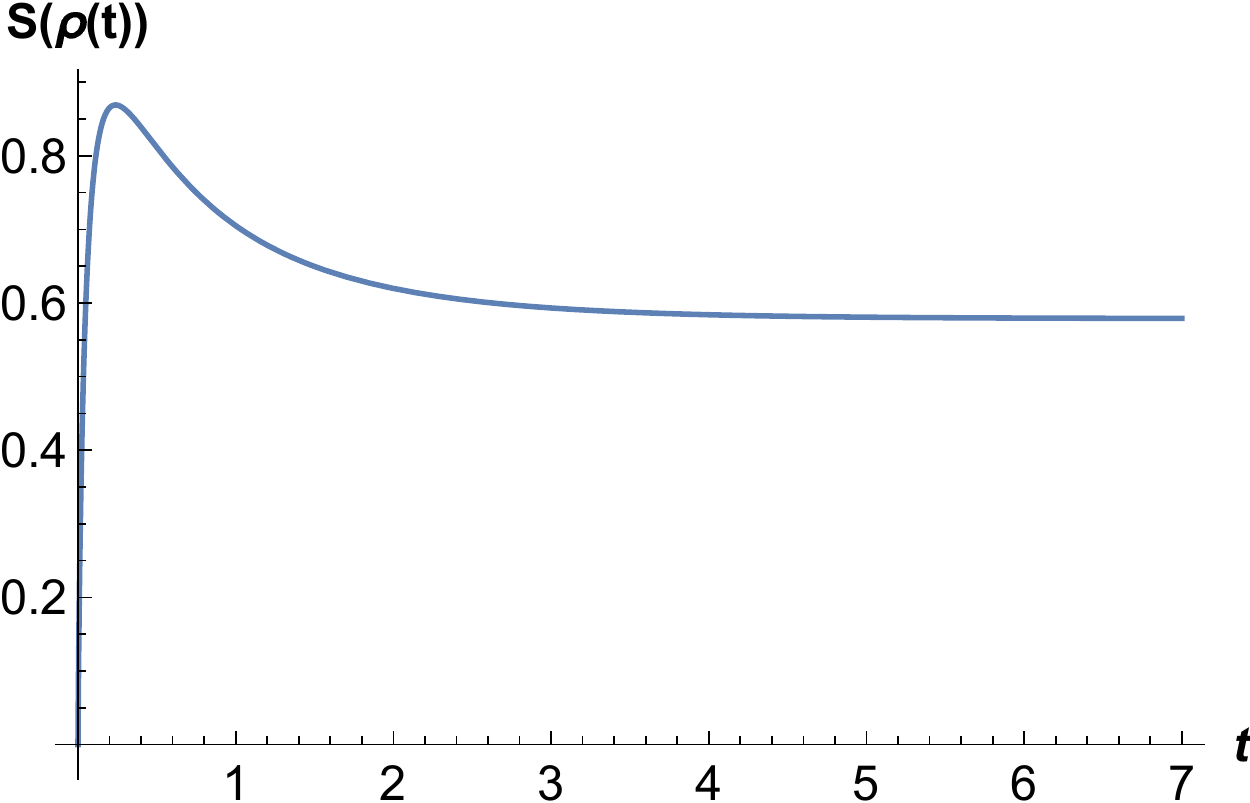}
	\caption{Entropy of the state $\rho(t)$ evolving according to
		(\ref{n_evol_nonzeroL}) with the initial condition and 
		parameters the same as in Fig.~\ref{fig8} 
		($\vecg{\xi}=(1/\sqrt{3},-1/\sqrt{3},-1/\sqrt{3})$,
		$g=-0.5$, $\omega=13$, $l=2.5$).}
	\label{fig9}
\end{figure}

Finally, we will discuss an example with nonzero Lindblad operators. 
For simplicity let us assume that only one of them is nonzero, i.e.,
$L_1=L$, $L_i=0$ for $i>1$.
In this case the nonlinear GKSL equation takes the form
\begin{equation}
 \dot{\rho} = -i[H,\rho] +\{G,\rho\}+
 L\rho L^\dagger - \rho \tr[\rho(2G+L^\dagger L)].
\end{equation}
Now, we consider an example of the Kraus form of the qubit evolution
with the following choice of generators:
\begin{equation}
G=-\kappa I+\tfrac{g}{2}\sigma_3, \quad
H=\tfrac{\omega}{2}\sigma_3,\quad
L=\tfrac{l}{2}(\sigma_1+i\sigma_3)
\end{equation}
with the initial condition given in Eq.~(\ref{initial_qubit}).
The nonlinear GKSL equation implies in this case that
\begin{align}
\dot{n}_3(t) & = -\Big(g-\tfrac{l^2}{2}\Big)(n_3(t))^2
 - l^2 n_3(t) + \Big(g+\tfrac{l^2}{2}\Big),\\
 \dot{n}_+(t) & = 
 \Big[ \Big(i\omega-\tfrac{l^2}{2}\Big) - 
 \Big(g-\tfrac{l^2}{2}\Big)n_3(t) \Big]n_+(t),
\end{align}
where $n_+=n_1+in_2$.
Taking into account the initial condition and integrating the system we
finally get
\begin{subequations}
\begin{align}
n_3(t)  & =
\frac{(\bar{l}+\xi_3)-(1-\xi_3)\bar{l}e^{-2gt}}{
	(\bar{l}+\xi_3)+(1-\xi_3)e^{-2gt}},\\
n_+(t) & = 
\frac{\xi_+(\bar{l}+1)e^{(i\omega-g)t}}{
	(\bar{l}+\xi_3)+(1-\xi_3)e^{-2gt}},
\end{align}%
\label{n_evol_nonzeroL}%
\end{subequations}%
where $\bar{l}=\frac{2g+l^2}{2g-l^2}$.
Notice that $g\le0$ implies $-1\le\bar{l}\le1$.
Moreover, for $0<g$: $n_3(\infty)=1$, $n_+(\infty)=0$ so we obtain the
pure state while for $0>g$: $n_3(\infty)=-\bar{l}$, $n_+(\infty)=0$ so
we get in general the mixed state.
In Fig.~\ref{fig8} we present an exemplary evolution of a Bloch vector
in this case.
In Fig.~\ref{fig9} we present the behavior of the von Neumann entropy
$S(\rho(t))$ under the evolution (\ref{n_evol_nonzeroL}), we take 
the same initial condition and parameters as in Fig.~\ref{fig8}.

We can also find an explicit form of Kraus operators in this case.
We have
\begin{equation}
K_0(t) = e^{-\kappa t} e^{\frac{1}{2}t(g-i\omega)\sigma_3}
 = e^{-\kappa t} \begin{pmatrix}
 e^{\frac{1}{2}(g-i\omega)t} & 0 \\
 0 & e^{-\frac{1}{2}(g-i\omega)t}
 \end{pmatrix},
\end{equation}
and
\begin{equation}
K_1(t)  = \frac{1}{2} e^{-\kappa t} e^{-\frac{1}{2}gt} \sqrt{e^{l^2t}-1}
 (\sigma_1+i\sigma_2).
\end{equation}
Furthermore
\begin{equation}
F(t) = e^{-2\kappa t}
\begin{pmatrix}
e^{gt} & 0 \\ 0 & e^{(l^2-g)t}
\end{pmatrix},
\end{equation}
so we arrive at the same form of $\vec{n}(t)$ as the result of the
solution of the nonlinear GKSL equation.

\subsection{The Jaynes--Cummings model}

In this section we show that the standard Jaynes--Cummings
model \cite{JC1963_JCM-original,SK1993_JCM}
describing the interaction of a two-level atom with a single mode of 
the electromagnetic (EM) field can be also generalized along the lines 
discussed in the paper.
The standard Jaynes--Cummings Hamiltonian describing the system
atom + EM field has the form
\begin{equation}
	H_{JC} = \omega_f \hat{a}^\dagger \hat{a} +
	\tfrac{\omega_a}{2} \sigma_3 +
	\tfrac{g}{2}(\hat{a}\sigma_+ + \hat{a}^\dagger\sigma_-).
	\label{JCM_hamiltoniam_full}
\end{equation}
The creation and annihilation operators $\hat{a}^\dagger$, $\hat{a}$
fulfill the standard bosonic commutation relation.

In order to obtain a nonlinear generalization we replace $H_{JC}$
with 
\begin{equation}
	H+iG = \big(  \omega_f \hat{a}^\dagger \hat{a} +
	\tfrac{\omega_a}{2} \sigma_3\big) +
	i \big(\tfrac{g}{2}(\hat{a}\sigma_+ + \hat{a}^\dagger\sigma_-)\big).
\end{equation}
Let us notice that subspaces
\begin{equation}
	\mathcal{H}^{(n)} = 
	\text{Span}\,\{ \ket{n}\otimes
	\begin{pmatrix}
		1 \\ 0
	\end{pmatrix}
	,\ket{n+1}\otimes
	\begin{pmatrix}
		0 \\ 1
	\end{pmatrix} \}
\end{equation}
are invariant under the action of both $H$ and $iG$.
Therefore, $H+iG$ can be written as the following direct sum:
\begin{equation}
	H+iG = \bigoplus_{n} H^{(n)},
	\label{hamiltonian_subspaces}
\end{equation}
where
\begin{equation}
	H^{(n)} = \omega_f(n+\tfrac{1}{2}) I + \tfrac{\omega}{2}\sigma_3
	+ i \tfrac{g}{2}\sqrt{n+1}\sigma_1.
	\label{JC_Hn}
\end{equation}
Furthermore, assuming that the initial full density matrix can be written in 
a similar way
\begin{equation}
	\rho(0) = \bigoplus_n \lambda_n(0) \rho_{(n)}(0),
	\label{rho_init_subsapces}
\end{equation}
where the conditions $\tr[\rho(0)]=1$, $\tr[\rho_n(0)]=1$ imply that
$\sum_n \lambda_n=1$.
Now, according to Eq.~(\ref{rho_t}), 
during the evolution governed by 
(\ref{hamiltonian_subspaces}) the density matrix
(\ref{rho_init_subsapces}) evolves to
\begin{equation}
	\rho(t) = \frac{e^{(G-iH)t} \rho(0) e^{(G+iH)t}}{\tr[e^{(G-iH)t} \rho(0) e^{(G+iH)t}]} 
	= \bigoplus_n \lambda_n(t) \rho_{(n)}(t),
	\label{JC_rhot}
\end{equation}
where
\begin{align}
	\rho_{(n)}(t) & =  \frac{e^{-itH^{(n)}} \rho_{(n)}(0) e^{itH^{(n)\dagger}}}{\tr[e^{-itH^{(n)}} \rho_{(n)}(0) e^{itH^{(n)\dagger}}]}, 
	\label{JC_rhont}\\
	\lambda_n(t) & = \lambda_n(0) \frac{\tr[e^{-itH^{(n)}} \rho_{(n)}(0) e^{itH^{(n)\dagger}}]}{\tr[e^{(G-iH)t} \rho(0) e^{(G+iH)t}]}.
\end{align}
Notice that choosing $\lambda_0=1$, $\lambda_i=0$ for $i>0$
we obtain the model considered in the subsection \ref{subsec:Lzero}.
From Eqs.~(\ref{JC_rhot}--\ref{JC_rhont}) it follows that in each subspace 
$\mathcal{H}^{(n)}$ the time evolution generator has the form (\ref{JC_Hn}), 
i.e., 
the corresponding vectors  $\vecg{\omega}_n$ and $\vec{g}_n$ are of the form 
$\vecg{\omega}_n=\omega(0,0,1)$ and $\vec{g}_n=g(\sqrt{n+1},0,0)$ 
so $\vecg{\omega}_n\cdot \vec{g}_n=0$. 
Thus, according to Eqs.~(\ref{n_asymptotic_ii}--\ref{n_asymptotic_iii}) 
if $g>\omega$ the oscillations are damped and 
in each subspace states evolve to appropriate stationary states. 
However, if $\omega\ge g$ then for lower $n$ satisfying $\omega\ge g\sqrt{n+1}$
the states are oscillating but for $g\sqrt{n+1}>\omega$
are evolving to stationary states.

\section{Applications}
\label{sec:applications}

In this section we present two applications of quasi-linear evolutions: 
relativistic Dirac qubit in electromagnetic field and evolution of the flavor 
state of solar neutrinos propagating through Sun in their way to Earth. 
In both cases we utilize the nonlinear differential evolution equation (\ref{nonlinvN_rho})  
generalizing the von Neumann equation. 
In the former case we assume time independence of the evolution  generators 
so we can use the global solution (\ref{nt-general-a-b}). 
In the latter case generators must be time(distant)-dependent which causes 
more sophisticated and interesting dynamics but solutions are usually achievable 
with use of numerical calculations only.

\subsection{Evolution of the relativistic qubit carrying by a Dirac 
	particle in electromagnetic field}
\label{sec:relativistic-qubit-EM}

Let us consider an evolution of a spin $1/2$ quantum charged particle 
(e.g. electron) which motion is generated by the electromagnetic  tensor  
$F^{\mu\nu}$.
We assume that state of the particle is determined by 
its four-momentum $p^\mu$ and the spin density matrix heaving in the spin 
basis standard form $\rho(\vecg{\xi})=\tfrac{1}{2}(I+\vecg{\xi}\cdot\vecg{\sigma})$,
where $\vecg{\xi}$ is the particle polarization vector, so $\vecg{\xi}^2\le1$.
For a sharp momentum particle it is possible to find this matrix in the manifestly
Lorentz-covariant spinorial basis \cite{cab_CR2005}
\begin{equation}
\Theta(p,w) = \tfrac{mc}{4p^0} 
(I+\tfrac{p_\mu}{mc}\gamma^\mu)
(I-2\gamma^5 \tfrac{w_\nu}{mc}\gamma^\nu) \gamma^0
\end{equation}
and $\tfrac{p^0}{mc}\Theta(p,w)\gamma^0$ is known in the literature 
\cite{cab_BLP1968} as the covariant spin density matrix for a Dirac particle. 
Here $\gamma^\mu$ are the gamma matrices  while $w^\mu$ is the counterpart 
of the Pauli-Lubanski four-vector, called polarization pseudo-vector. 
It is defined by the relations
$w^0(p,\vecg{\xi}) = \tfrac{1}{2} \vec{p}\cdot\vecg{\xi}$,
$\vec{w}(p,\vecg{\xi}) = \tfrac{1}{2}(mc\vecg{\xi} +
\frac{\vec{p}(\vec{p}\cdot\vecg{\xi})}{p^0+mc})$
\cite{cab_BLT1969}. Notice, that $\vecg{\xi}$ can be expressed by $p$ and $w$:
\begin{equation}
\vecg{\xi} = \frac{2}{mc} \Big(
\vec{w} - w^0 \frac{\vec{p}}{p^0+mc}
\Big).
\end{equation}
The four-vectors $p_\mu$ and $w^\mu$ are Minkowski-orthogonal, i.e., 
$p_\mu w^\mu=0$.  Moreover
$p_\mu p^\mu = (mc)^2$, $w_\mu w^\mu = - (\tfrac{mc}{2})^2 \vecg{\xi}^2$.
Under the bi-spinor representation of the Lorentz group $\Theta$ 
transforms according to
\begin{equation}
\Theta^\prime = \frac{S \Theta S^\dagger}{\tr(S \Theta S^\dagger)}
\end{equation}
with
$S= \exp(i\omega_{\mu\nu}J^{\mu\nu})$ and 
$J^{\mu\nu} = \tfrac{i}{4}[\gamma^\mu, \gamma^\nu]$.
In the following we will use the Weyl representation of the gamma matrices 
(see Appendix \ref{sec:appB}).  It is easy to prove that $\Theta$ is Hermitian, 
non-negative definite and $\tr\Theta=1$ as well as it is related to the matrix
$\rho(\vecg{\xi})$ by the formulas
\begin{align}
\rho(\vecg{\xi}) & = \overline{v(p)}\Theta(p,w)v(p),\\
\Theta(p,w) & = v(p) \rho(\vecg{\xi}) \overline{v(p)},
\end{align}
where $v(p)$ satisfies the Dirac equation and intertwines spinorial and spin bases; 
its explicit form in the Weyl representation is given in Appendix \ref{sec:appB}.

Now, we define the evolution of $\Theta$, satisfying the nonlinear von Neumann equation 
(\ref{nonlinvN_rho}), by choosing its generator as follows
\begin{equation}
H+iG = \frac{\mu_B}{\hbar} F_{\mu\nu} J^{\mu\nu}
= i \frac{\mu_B}{c\hbar} 
\begin{pmatrix}
\vec{F}\cdot\vecg{\sigma} & 0 \\
0 & - \vec{F}^*\cdot\vecg{\sigma}
\end{pmatrix},
\end{equation}
where $\mu_B=\tfrac{e\hbar}{2m}$ is the Bohr magneton for a particle with 
charge $e$. The complex vector
\begin{equation}
\vec{F} = \vec{E} + ic \vec{B}
\end{equation}
is known as the Riemann-Silberstein vector related to a self-dual tensor 
while $\vec{E}$, $\vec{B}$ are the electric and magnetic fields, respectively.
Note that
\begin{align}
H & = - \frac{\mu_B}{\hbar} 
\begin{pmatrix}
\vec{B}\cdot\vecg{\sigma} & 0\\
0 & \vec{B}\cdot\vecg{\sigma}
\end{pmatrix},\\
G & = \frac{\mu_B}{c\hbar} 
\begin{pmatrix}
	\vec{E}\cdot\vecg{\sigma} & 0\\
	0 & -\vec{E}\cdot\vecg{\sigma}
\end{pmatrix},
\end{align}
so $H$ coincides with the Hamiltonian (divided by $\hbar$) 
twisting the spin and in the case of a constant $\vec{B}$.

\begin{figure*}
	\subfigure[]{%
		\label{10a}
		\includegraphics[width=.45\textwidth]{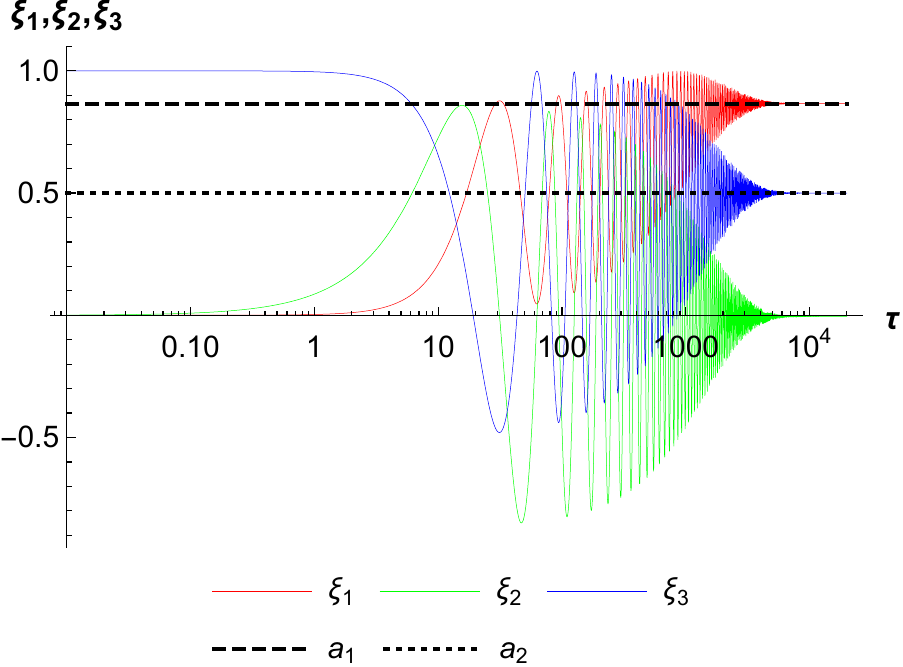}}
	\hspace{0.03\textwidth}
	\subfigure[]{%
		\label{10b}
		\includegraphics[width=.45\textwidth]{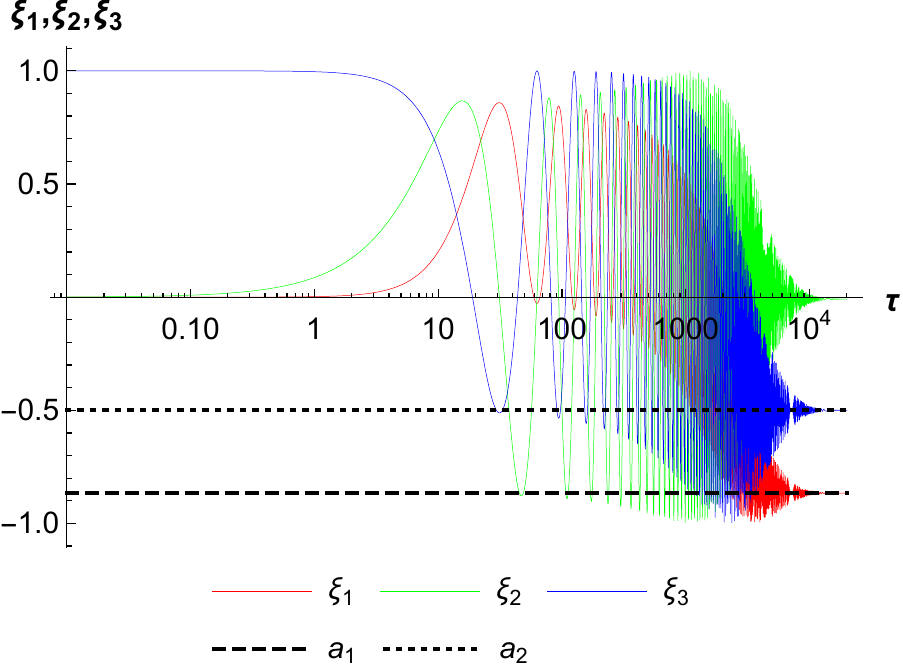}}
	\caption{The behavior of components of the polarization vector $\vecg{\xi}$ during 
		the evolution of a relativistic qubit (compare Sec.~\ref{sec:relativistic-qubit-EM})
		in the electromagnetic field.
		The fields $\vec{B}$ and $\vec{E}$ are chosen in the plane $xz$ with 
		the polar angles $\pi/3$ and $\pi/2$ (for Fig.~\ref{10a}) and 
		$\pi/3$ and $\pi$ (for Fig.~\ref{10b}). 
		Notice, that this state evolves to the stationary states which are
		very close to the Hamiltonian $H$ eigenstates Bloch vectors
		$\pm (\sin\tfrac{\pi}{3},0,\cos\tfrac{\pi}{3})$.
		The exact eigenstates are achieved for parallel (antiparallel)
		configurations of $\vec{B}$ and $\vec{E}$. 
		Here strength of the fields $\vec{B}$ and $\vec{E}$ is of 
        the order of $5\times10^{-13}$~T ($\omega\sim0.1$) and 
        $2\times10^{-13}$~V/m ($g\sim0.001$), respectively, 
        while the initial value of the Bloch  vector 
		$\xi_0 = (0, 0, 1)$. $\tau$-axis in logarithmic scale.}
	\label{fig10}
\end{figure*}

Assuming that  the electromagnetic field is time independent we obtain the 
result of integration of the von Neumann equation in the form (\ref{rho_t}). 
More specifically, the evolution operator takes the form
\begin{equation}
K(\tau) = e^{-i\tau \frac{\mu_B}{\hbar} F_{\mu\nu} J^{\mu\nu}} =
\begin{pmatrix}
e^{\tau\frac{\mu_B}{c\hbar} \vec{F}\cdot\vecg{\sigma}} & 0 \\
0 & e^{-\tau\frac{\mu_B}{c\hbar} \vec{F}^*\cdot\vecg{\sigma}}
\end{pmatrix},
\label{application-K-t}
\end{equation}
where we choose the proper time $\tau$ as the dynamical parameter to guarantee 
the Lorentz covariance of our formalism. 
Taking into account that  $\gamma^0 K^\dagger(\tau) = K^{-1}(\tau)\gamma^0$, 
and choosing the rest frame values of the four-momentum and the polarization 
pseudo-vector as the initial conditions, i.e., 
$p(0)= mc(1;0,0,0)$ so $w(0)=mc(0;\vecg{\xi}_0)$, we obtain that
\begin{equation}
\Theta(\tau) = \frac{K(\tau) \Theta(p(0),w(0)) K^\dagger(\tau)}
{\tr[K(\tau) \Theta(p(0),w(0)) K^\dagger(\tau)]},
\end{equation}
and
\begin{equation}
\rho(\vecg{\xi}(\tau)) = \frac{e^{\tau\frac{\mu_B}{c\hbar}
\vec{F}\cdot\vecg{\sigma}} \rho(\vecg{\xi}_0) 
e^{\tau\frac{\mu_B}{c\hbar} \vec{F}^*\cdot\vecg{\sigma}}}{
\tr\big[ e^{\tau\frac{\mu_B}{c\hbar}
	\vec{F}\cdot\vecg{\sigma}} \rho(\vecg{\xi}_0) 
e^{\tau\frac{\mu_B}{c\hbar} \vec{F}^*\cdot\vecg{\sigma}} \big]},
\label{applicatio-rho-t}
\end{equation}
while $p(\tau)$ as well as $w(\tau) $ can be calculated by means of the following 
equations
\begin{align}
\sigma_\mu p^\mu(\tau) & = 
e^{\tau\frac{\mu_B}{c\hbar}
	\vec{F}\cdot\vecg{\sigma}}
\sigma_\mu q^\mu e^{\tau\frac{\mu_B}{c\hbar}
	\vec{F}^*\cdot\vecg{\sigma}},
\label{application-sigma-p-tau}\\
\sigma_\mu w^\mu(\tau) & = 
e^{\tau\frac{\mu_B}{c\hbar}
	\vec{F}\cdot\vecg{\sigma}}
\sigma_\mu \vartheta^\mu e^{\tau\frac{\mu_B}{c\hbar}
	\vec{F}^*\cdot\vecg{\sigma}}.
\label{application-sigma-w-tau}
\end{align}
The solution of (\ref{applicatio-rho-t})  is given in (\ref{nt-general-a-b}) under appropriate
identification of $\vec{F}$ and $\vecg{\xi}$ with $\vecg{\omega}$, $\vec{g}$, and 
$\vec{n}$. 
On the other hand, from (\ref{application-K-t}) with use of the explicit form 
of the gamma matrices in the Weyl form (Appendix \ref{sec:appB}) we see 
that  Eqs.~(\ref{application-sigma-p-tau}-\ref{application-sigma-w-tau}) 
imply the following differential equations for the four-momentum $p(\tau)$ and 
the polarization four-vector $w(\tau)$:
\begin{align}
\frac{d p^\mu}{d\tau} & = \frac{e}{m} F^\mu_{\phantom{\mu}\nu} p^\nu,
\label{application-diff-eq-p}\\
\frac{d w^\mu}{d\tau} & = \frac{e}{m} F^\mu_{\phantom{\mu}\nu} w^\nu,
\label{application-diff-eq-w}
\end{align}
Equation (\ref{application-diff-eq-p})  is the well known covariant form of 
the classical charged particle momentum evolution under action of the Lorentz 
force and its  energy change \cite{Jackson1999} 
while Eq.~(\ref{application-diff-eq-w}) is equivalent to the evolution equation 
for the classical pseudo-spin vector of a particle with the gyromagnetic 
ratio value equal to 2 (Bargmann-Michel-Telegdi (BMT) equation \cite{cab_BMT1959}).
Summarizing, we demonstrated the semi-quantum analog of the BMT formalism 
where space-time coordinates are treated classically. Here the spin state is quantum. 
Indeed, heaving $\vecg{\xi}(\tau)$ and $p(\tau)$ we have determined the density 
matrix $\Theta$ so we are able to calculate probabilities of measurements and 
evolution of averages of relativistic observables for the Dirac particle being the 
carrier particle of the relativistic qubit. 
We demonstrate evolution of the spin state of electron moving in the constant 
electromagnetic field in Fig~\ref{fig10}. 
Notice, that this state evolves to the state which is very close to 
stationary state of the Hamiltonian $H$. 
The exact stationary states of $H$ are achieved for parallel (antiparallel) 
configurations of $\vec{E}$ and $\vec{B}$.

\subsection{Evolution of the flavor  quantum state of solar neutrino}
\label{sec:neutrino-flavor}

As is well known, neutrino oscillates in the leptonic flavor space 
(space of fractional leptonic charges), i.e., it evolves between electric 
muonic and taonic neutrino flavors. 
However, neutrinos produced in the core of Sun undergo interaction 
with the dense solar electron plasma and, as consequence, their flavor oscillations 
are modified when neutrinos propagate through the Sun in their way to detectors 
on Earth. 
This is known as the Mikheyev-Smirnov-Wolfenstein (MSW) mechanism 
based on  the Wolfenstein equation and form of the solar electron density function 
\cite{MS2016-Solar-neutrinos}. 
This process is dominated by the electron and muon neutrinos. Therefore, the effective
Hamiltonian in the flavor space of electronic and muonic neutrinos is of the form
\begin{align}
H & = \frac{\epsilon}{2}
\begin{pmatrix}
- \frac{\Delta m^2 \cos(2\theta)}{2 E} + V(L) & \frac{\Delta m^2 \sin(2\theta)}{2 E}\\
\frac{\Delta m^2 \sin(2\theta)}{2 E} & \frac{\Delta m^2 \cos(2\theta)}{2 E} - V(L)
\end{pmatrix}\nonumber\\
& \equiv \frac{\epsilon}{2} \vecg{\omega}(L) \cdot \vecg{\sigma}.
\end{align}
Here $E$ is the neutrino energy given in GeV, $\theta=\theta_{12}=0.59$~rad
is the mixing angle between mass 1 and 2 Bloch vectors for eigenstates $\nu_1$ and $\nu_2$,
$\Delta m^2 = m_2^2 - m_1^2 \approxeq 8 \times 10^{-5}$~eV$^2$, 
\begin{widetext}
\begin{equation}
V(L) = \sqrt{2} G_F \rho(L) \approxeq
\begin{cases}
0.012 (519 \tfrac{L^4}{R_S^4} - 1630 \tfrac{L^3}{R_S^3}
+1844 \tfrac{L^2}{R_S^2} - 889 \tfrac{L}{R_S} +154.910686)
& \text{for}\quad 0<L\le 365767, \\
0 & \text{for}\quad 365767<L,
\end{cases}
\end{equation}
\end{widetext}
(where Sun radius $R_S=695700$~km)
approximates the effective potential (given in neV) generated by the solar electron 
plasma density $\rho(L)$ and $L$ is the radial distance from the center of Sun 
given in kilometers; $G_F$ is the Fermi constant while $\epsilon=5.08$ is 
the conversion factor. 
The Wolfenstein equation is simply the Schr\"odinger equation with time 
$t$ replaced by $L=ct$, i.e., it describes evolution of the neutrino flavor 
state in the neutrino travel from center of Sun to Earth.

In the paper \cite{RCib2021_in_prep}
one demonstrates another point of view to this problem suggesting 
that the change of the neutrino state inside of the dense matter of Sun can 
be forced by a sort of damping of the neutrino flavor oscillations and the 
full evolution generator should be of the form  $H+i G$ with
\begin{equation}
G = \tfrac{\epsilon}{2} \vec{g}(L)\cdot\vecg{\sigma},
\end{equation}
and $|\vec{g}(L)|=\sqrt{2} G_F \rho(L)$ while
the evolution equation have the form of the quasi-linear Schr\"odinger 
equation (\ref{nonlinvN_psi}):
\begin{equation}
i \frac{d}{dL} \psi = \frac{\epsilon}{2}
\big(
\vecg{\omega}(L)\cdot\vecg{\sigma} 
+ i \vec{g}(L)\cdot\vecg{\sigma}
- i \vec{g}(L) \psi^\dagger \vecg{\sigma} \psi
\big) \psi,
\end{equation}
where
\begin{equation}
\psi = \begin{pmatrix}
	\nu_e \\ \nu_\mu
\end{pmatrix}
\quad \text{and}\quad
\psi^\dagger \psi = 1,
\end{equation}
and $\nu_e$, $\nu_\mu$ 
denotes the electronic and muonic  flavor states, respectively.

\begin{figure*}
	\subfigure[]{%
		\label{11a}
		\includegraphics[width=.45\textwidth]{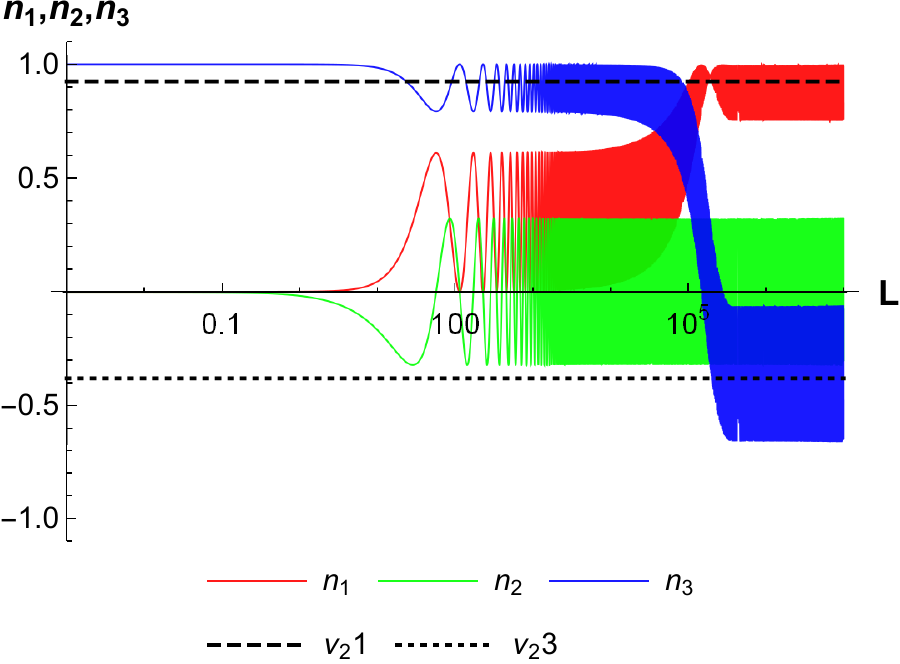}}
	\hspace{0.03\textwidth}
	\subfigure[]{%
		\label{11b}
		\includegraphics[width=.45\textwidth]{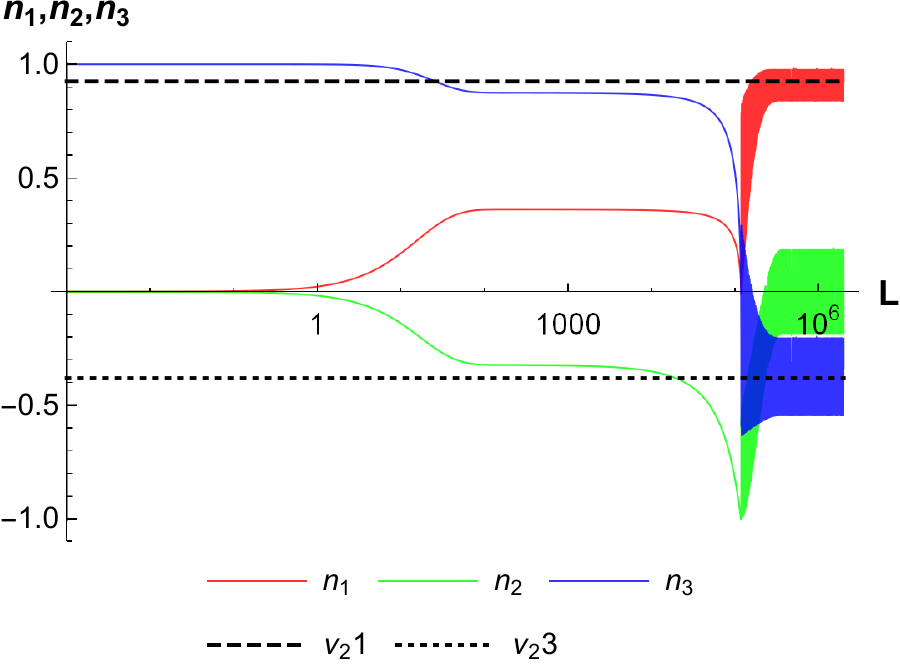}}
	\caption{The evolution of components of the Bloch vector describing the neutrino state.
	In Fig.~\ref{11a} we present standard MSW evolution 
	($V\not=0$, $\vec{g}=\vec{0}$),
	in Fig.~\ref{11b} we present evolution generated by $\vec{g}$ 
	($V=0$, $\vec{g}\not=\vec{0}$) for typical 10~MeV energy neutrino flux.
	For the details see Sec.~\ref{sec:neutrino-flavor}.
	$L$-axis in logarithmic scale, $\nu_22=0$.}
	\label{fig11}
\end{figure*}

In Fig.~\ref{fig11} we compare both  evolutions, standard MSW 
($V\not=0$, $\vec{g}=\vec{0}$) in Fig.~\ref{11a} and generated by $\vec{g}$ 
($V=0$, $\vec{g}\not=\vec{0}$) in Fig.~\ref{11b} for typical 10~MeV energy neutrino flux. 
Both approaches give the same probabilities of registration of electronic (muonic) neutrinos 
on Earth measured, e.g., in Super-Kamiokande detector or Sudbury Neutrino Observatory. 
However, it is very interesting that in the latter case, the resonant-like conversion 
to the final state holds in the neighborhood of the point of the structural instability 
of the evolution where infinitesimal change of any of the parameters changes 
drastically the character of the evolution \cite{AAIS1991}.
In the considered case it holds for $\vec{g}$ perpendicular to $\vecg{\omega}$ 
in the point $L_{in}$ of the trajectory where $|\vec{g}|=|\vecg{\omega}|$,  
i.e., when  $\sqrt{2} G_F \rho(L_{in}) =  \frac{\Delta m^2}{2E}$ 
so $L_{in} \approxeq 117600$~km (see Fig.~\ref{11b}). 
Before this point we observe the strong flavor oscillation damping while after 
conversion the neutrino state stabilizes as oscillations close to 
the Bloch vector of $\nu_2$ mass neutrino stationary state. 
In contrast, the MSW model does not  show any significant disappearance of 
oscillations also in neighborhood of MSW critical (resonant) point 
$L_c \approxeq 191300$~km identified by the condition 
$\sqrt{2} G_F \rho(L_{c}) =  \frac{\Delta m^2}{2E} \cos(2\theta)$.
An exhaustive discussion of the above questions together with experimental implications
is given in the paper prepared for publication \cite{RCib2021_in_prep}.

\section{Discussion and conclusions}
\label{sec:Conclusion}

In this paper we have discussed a generalization of the notion 
of the quantum operations and the quantum time evolution. 
The generalization depends on extending the linearity of quantum operations 
to the quasi-linearity condition.
This condition is motivated by the appearance of such operations in a 
``hidden''  form (e.g. selective measurement \cite{Kraus1983})
in quantum formalism.
On the other hand, convex quasi-linearity guarantees absence of
superluminal communication.
We have identified a natural class of operations satisfying this condition. 
Moreover, we have generalized the GKSL master equation for 
quasi-linear evolutions.
As an example we have considered nonlinear qubit evolution.
It is worth to stress
that some of these qubit evolutions were discussed
independently in the different
context of non-Hermitian quantum mechanics
\cite{BG2012_non-Hermitian-gain-loss,%
SZ2013_non-Hermitian_q_dyn_two-level,GCKM_su11_Hamiltonian}.
We also discussed an appropriate modification of the Jaynes--Cummings model
\cite{JC1963_JCM-original,SK1993_JCM} describing the interaction
of a two-level atom with a single mode of the electromagnetic field.
In general, the nonlinear time development of qubit is related to an interaction 
of the quantum system with an environment. 
Depending on the interrelation with environment the evolution
of qubit can demonstrate different time dependence
but mostly it can be interpreted as a damping of oscillations
together with gain and loss of energy (Sec.~\ref{sec:time-dependent-g}). 
This is especially evident in presented physical applications of the introduced formalism:
evolution of spin of a relativistic Dirac particle in external electromagnetic field 
and evolution of the flavor state of solar neutrinos propagating through the Sun. 
In both cases we observe damping of oscillations of the state and its asymptotic 
evolution to a stationary state. Moreover, the later case offers a new physical 
interpretation of the process of transmutation of neutrinos belonging to different 
flavor generations inside Sun.
  
To avoid misinterpretations of the introduced formalism we stress that 
the generator $G$ in the master equation (\ref{GKSL_nonlinear_general}) describes, 
similarly as the Lindblad generators, influence of an environment on 
a quantum system governed on the free level by the Hamiltonian $H$. 
This means that the $H+i G$ appearing in global solutions 
(\ref{Kt_vanishing_Lindblad-1}), (\ref{Kt_no_Lindblads_gen}) of 
the nonlinear von Neumann equation (\ref{nonlinvN_rho}) or 
in the nonlinear Schr\"odinger equation (\ref{nonlinvN_psi}) is 
the evolution generator, however, in general, it cannot be interpreted as a non-Hermitian
$PT$-symmetric Hamiltonian. Notice, that the spectrum of $H+i G$ 
is in general complex while $PT$-symmetric Hamiltonians have real spectra,
i.e., $PT$-symmetric Hamiltonians form a subset of the considered
set of generators $H+iG$
while the corresponding trace-preserving, $PT$-symmetric evolutions belong to 
the introduced by us class of quasi-linear, so admissible, evolutions.
This fact supports their physical acceptance because it shows that they do not lead to
arbitrary fast communication.

Let us stress here that our goal was to introduce nonlinear evolution
with minimal changes in the rest of quantum formalism. 
In our approach we do not change Born rule or projection postulate
as it takes place in other nonlinear extensions of quantum mechanics
(\cite{CD2002_Correl-exp-nonlin_PLA,HCh2017_Born_NLQM}).
But of course
nonlinearity of evolution equations has implications---the superposition 
principle is broken during the nonlinear evolution.

\begin{acknowledgments}
We thank Dariusz Chru\'sci\'nski, Jacek Ciborowski and Pawe{\l} Horodecki 
for interesting discussions.
This work has been supported by the Polish National Science Centre
under the contract 2014/15/B/ST2/00117 and by the University of Lodz.
\end{acknowledgments}

\appendix
\section{Derivation of Eq.~(\ref{GKSL_nonlinear_general})}
\label{sec:appA}

To derive the infinitesimal form of the global time evolution (\ref{nonlin_dyn_global}) we proceed as follows:
First, we put
\begin{equation}
\Phi_{t+\delta t}(\rho_{0}) \approx 
\Phi_t(\rho(0)) + \delta t \dot{\Phi}_t(\rho_0) = 
\rho(t) + \delta t \dot{\rho}(t),
\label{app_A1}
\end{equation}
where $\rho(t)=\Phi_t(\rho_{0})$.
On the other hand, using the composition law of the map $\Phi_t$
we obtain
\begin{equation}
\Phi_{t+\delta t}(\rho_{0}) = 
\Phi_{\delta t}\big( \Phi_t(\rho_0) \big) =
\Phi_{\delta t}(\rho(t)).
\end{equation}
Next, with the help of the global form (\ref{nonlin_dyn_global})
we get
\begin{equation}
\Phi_{\delta t}(\rho(t)) = \frac{\sum_{\alpha=0}^{\alpha_{max}} K_\alpha(\delta t) \rho(t) 
K_{\alpha}^{\dagger}(\delta t)}{\sum_{\alpha=0}^{\alpha_{max}} 
\tr\big( K_{\alpha}^{\dagger}(\delta t) K_\alpha(\delta t) \rho(t) \big)}.
\label{Phi_delta_t}
\end{equation}
The initial condition implies
\begin{equation}
K_0(0) = I,\quad K_\alpha(0)=0\,\, \text{for}\,\, \alpha\ge1.
\end{equation}
Next, we expect that in the expansion of the sum 
$\sum_{\alpha=0}^{\alpha_{max}}
 K_\alpha(\delta t) \rho(t) K_{\alpha}^{\dagger}(\delta t)$
the first correction is of the order $\delta t$. Therefore, we put
\begin{equation}
K_0(\delta t) = I+ \delta t \kappa, \quad
K_\alpha(\delta t) = \sqrt{\delta t} L_\alpha\,\, \text{for}\,\, \alpha\ge1,
\end{equation}
where $\kappa$ and $L_\alpha$ are linear operators.
Now, the nominator of Eq.~(\ref{Phi_delta_t}) takes the form
\begin{multline}
\sum_{\alpha=0}^{\alpha_{max}}
K_\alpha(\delta t) \rho(t) K_{\alpha}^{\dagger}(\delta t)
=
\rho(t) + \delta t \big(\kappa \rho(t) + \rho(t) \kappa^\dagger\big)\\
+ \delta t \sum_{\alpha=1}^{\alpha_{max}}
L_\alpha \rho(t) L_{\alpha}^{\dagger}.
\label{app_nominator}
\end{multline}
Using this equation we obtain
\begin{multline}
\tr\Big(\sum_{\alpha=0}^{\alpha_{max}}
K_\alpha(\delta t) \rho(t) K_{\alpha}^{\dagger}(\delta t)\Big)
= \\
1 + \delta t \tr\Big[ \rho(t)\Big( \kappa + \kappa^\dagger +
\sum_{\alpha=1}^{\alpha_{max}}
L_{\alpha}^{\dagger} L_\alpha
 \Big) \Big].
\end{multline}
Next, with the help of the expansion $(1+\delta x)^{-1}\approx 1- \delta x$,
the inverse of the denominator of Eq.~(\ref{Phi_delta_t}) takes the form
\begin{equation}
1 - \delta t \tr\Big[ \rho(t)\Big( \kappa + \kappa^\dagger +
\sum_{\alpha=1}^{\alpha_{max}}
L_{\alpha}^{\dagger} L_\alpha
\Big) \Big].
\label{app_denominator}
\end{equation}
Now, multiplying the right hand side of
Eq.~(\ref{app_nominator}) with 
Eq.~(\ref{app_denominator}), leaving the terms of order
up to $\delta t$, inserting $\kappa=G-iH$, 
and equating the result with (\ref{app_A1})
we obtain equation (\ref{GKSL_nonlinear_general}).

\section{Gamma matrices and other formulas}
\label{sec:appB}
Weyl representation of the gamma matrices
\begin{equation}
\gamma^0 = \begin{pmatrix}
	0 & I \\ I & 0
\end{pmatrix},\quad 
\gamma^k = \begin{pmatrix}
0 & \sigma_k \\ -\sigma_k & 0
\end{pmatrix},\quad
\gamma^5 = \begin{pmatrix}
-I & 0 \\ 0 & I
\end{pmatrix}.
\end{equation}

The explicit form of $v(p)$ is the following:
\begin{equation}
v(p) = \frac{1}{2\sqrt{1+\frac{p^0}{mc}}}
\begin{pmatrix}
I + \frac{1}{mc}(p^0 I - \vec{p}\cdot \vecg{\sigma})\\[1mm]
I + \frac{1}{mc}(p^0 I + \vec{p}\cdot \vecg{\sigma})
\end{pmatrix},
\end{equation}
moreover, it holds
\begin{equation}
\overline{v(p)} = v(p)^\dagger \gamma^0. 
\end{equation}


\end{document}